\documentclass[12pt,preprint,a4]{iopart}
\usepackage[pdftex]{graphicx}
\usepackage[utf8]{inputenc}
\usepackage[english]{babel}
\usepackage{ifthen}
\usepackage{subfigure}
\usepackage{epsfig}
\usepackage{enumerate}
\usepackage{array}
\usepackage{graphicx}
\usepackage{booktabs}

\usepackage{siunitx}
\usepackage[normalem]{ulem}

\newcommand{\waterone}{H$_3$O$^+$}
\newcommand{\watertwo}{(H$_2$O)$_2$H$^+$}
\newcommand{\waterthree}{(H$_2$O)$_3$H$^+$}
\newcommand{\waterfour}{(H$_2$O)$_4$H$^+$}

\usepackage[style=phys,biblabel=brackets,backend=biber,citestyle=numeric-comp,sorting=none]{biblatex}
\begin{document}

\title[IEDFs from the impact of a dielectric barrier discharge plasma jet on surfaces]{Ion energy distributions from the impact of an atmospheric dielectric barrier discharge plasma jet on surfaces} 

\author{Daniel Henze, Laura Chauvet, Achim von Keudell}
\address{Experimental Physics II - Reactive Plasmas, Ruhr-Universit\"at Bochum, D-44780 Bochum, Germany}

\date{\today}

\ead{Daniel.Henze@rub.de}
 
\begin{abstract}
The ion energy distribution functions (IEDF) have been measured for a helium atmospheric pressure dielectric barrier discharge jet expanding into the air and impacting a metal or ceramic surface. The plasma jet produces ionization waves as guided positive streamers that reach the surface. Molecular beam mass spectrometry (MBMS) with an energy filter has been used to monitor the IEDFs at a distance of 1.5 cm from the dielectric barrier discharge plasma jet exit. The species are sampled from the supersonic expanding helium beam passing into the MBMS through a 40 $\mu$m (metal) or a 50 $\mu$m (ceramic) diameter orifice. N$_2^+$, O$_2^+$, NO$^+$, O$_3^+$ and water cluster ions (H$_2$O)$_n$H$^+$ (n=1...4) are abundantly produced in the discharge. The analysis of the time-resolved IEDFs reveals that all ions are predominantly sampled at a reference energy $E_{\rm beam}$ when using the metallic orifice. This energy $E_{\rm beam}$ is determined by the seeding of the ions into the supersonic expanding helium beam into the MBMS. After the impact of the streamer, an afterglow of 10 $\mu$s is observed when ions are continuously sampled at an energy higher than $E_{\rm beam}$ by a few 0.1 eVs. This is resolved by postulating a positive space-charge region in front of a positively charged surface. The temporal sequence of the ion impact is consistent with reaction schemes in air plasmas, where O$_2^+$ and N$_2^+$ are created before the formation of NO and O$_3$, as well as larger water cluster ions.    
\end{abstract}

\maketitle

\section{Introduction}

Atmospheric-pressure plasma jets are an important application of plasmas for the treatment of surfaces, for the activation of liquids, or for biomedical applications \cite{tendero_atmospheric_2006,winter_atmospheric_2015,viegas_physics_2022,kondeti_longlived_2018,weltmann_atmosphericpressure_2010}. Depending on the plasma generation method, these plasma jets may produce either an effluent of radicals or a plasma beam that impacts a surface to be treated. Many of these plasmas are based on initiating a streamer \cite{nijdam_physics_2020} that propagates along a gas stream or dielectric tube up to several meters \cite{robert_experimental_2009} reaching very high propagation velocities \cite{Xiong_2010}. A typical dielectric barrier discharge plasma jet consists of a quartz capillary with two ring electrodes wrapped around it to generate an electric field which points along the gas flow so that an ionization wave is launched that propagates along the effluent. A typical excitation scheme uses voltages of several kV with a pulsed excitation from nanoseconds to $\mu$s. ICCD imaging is used to monitor these ionization waves, which had been coined so-called \textit{plasma bullets} that propagate as negative or positive guided streamers along the gas effluent \cite{mericam-bourdet_experimental_2009,jarrige_formation_2010,lu_atmosphericpressure_2012,Lu_2014,yan_gas_2017}. Streamer propagation along a gas stream has been modelled by several groups \cite{aleksandrov_simulation_1996,babaeva_ion_2011,babaeva_plasma_2019} which yielded results consistent with experimentally observed propagation velocities and electric field strengths \cite{vanderschans_electric_2017}. When these streamers impact a surface, the charge transport by the plasma affects the charge state at the surface depending on the dielectric constant of all the materials involved, as has been studied extensively \cite{klarenaar_how_2018,viegas_interaction_2020,guaitella_impingement_2015,slikboer_charge_2017}.

The ionization waves can be regarded as the head of a streamer (of either polarity according to the excitation source) that impacts the surface on an interaction time scale of a few nanoseconds. A high density of radicals and ions is created in front of the surface. Because the electric field at the streamer head is very large, one may speculate that high-energy ions can reach the surface at considerable energy within a mean free path length. Babaeva and Kushner studied such streamer impacts on various surfaces by computer simulation in a series of papers \cite{babaeva_ion_2011,babaeva_ion_2011a,babaeva_control_2013}. By assuming a flat surface as the target, they showed that the IEDFs of most ions exhibit kinetic energies of less than 1 eV, with a high-energy tail reaching only a few eV. The contribution of these higher-energy ions constitutes only a small fraction of the total ion flux, and these ions are only present in a very narrow time window of a few nanoseconds when the streamer head hits the surface. The energization of the ions is explained by them as follows: When the streamer head approaches the surface, the potential drop in the sheath in front of the surface is squeezed between the streamer head and surface, leading to a very large electric field and thus to a stronger acceleration within the mean free path of the ions. This squeezing of the sheath is a dynamic process that depends sensitively on the dielectric constant of the surface. For a high-dielectric-constant material and thus a high capacitance, the charging of the surface by the streamer leads only to a small voltage increase at the very surface so that the sheath voltage is preserved much longer, creating more energetic IEDFs than for a surface material with a small dielectric constant and a small capacitance. The plasma sheath and a dielectric surface layer on top of a grounded metal can be regarded as a capacitive divider. Based on this finding, Babaeva and Kushner analyzed the streamer impact on different materials, either on small particles representing a surface with a small capacitance (due to the small size of the particle), a flat thick dielectric layer, or a pore inside this dielectric. They found ion energies from a few eV for the case of streamer impact on small particles to energies up to 10...15 eV for the streamer impact on a thick dielectric, up to 200 eV for the case of a streamer penetrating a pore in the dielectric. However, said IEDFs from a streamer impacting a surface have never been measured.

The ion composition of atmospheric-pressure plasmas has been studied quantitatively using molecular beam mass spectrometry (MBMS) by Jiang and Bruggeman in the plasma afterglow \cite{jiang_absolute_2021} with an emphasis on the absolute quantification of the ion fluxes using a corona discharge as a calibration source. No analysis of the temporal structure and the energy spectra of the sampled ions had been performed. Ito et al. analysed the composition of ions generated by an air plasma \cite{ito_mass_2015}. A similar analysis has been performed by Oh et al. \cite{oh_investigating_2015}. The sampled ion compositions in these studies agree with the chemistry of such plasma in air \cite{naidis_production_2014}. At first, oxygen and nitrogen are ionized as well as dissociated. The reaction of O atoms or N atoms with N$_2$ or O$_2$, respectively, leads to the formation of NO. The reaction of O with O$_2$ leads to ozone formation in a three-body reaction. These reactive oxygen species may then react with NO to form NO$_2$ and larger nitric oxides. All of these reactions are sensitive to the electron temperature and density in the discharge. In particular, O$_3$ is swiftly destroyed in the presence of reactive species, so the formation is usually much higher in the plasma afterglow.

The ion contribution to the reactive species flux in atmospheric-pressure plasmas is usually considered to be of minor importance for many applications because their energies are low due to the collisional plasma sheaths, and the ions from a remote plasma source may not reach the treated surface. Nevertheless, it remains an open question whether energetic ions may interact with the surface in the highly dynamic environment of an impinging streamer. In that case, even a small fraction of ions might have a significant impact because their energy is sufficient to break molecular bonds and affect the surface material. 

The measurement of the ion energy distribution functions from an atmospheric pressure plasma using MBMS has been addressed by \cite{bruggeman_mass_2010} a while ago, showing that the diameter of the orifice is crucial for the determination of the IEDFs. In general, the diameter of the orifice needs to be smaller than the Debye length to avoid any plasma penetration into the MBMS. Otherwise, ionization also takes place inside the MBMS and very high energies might be measured due to the acceleration of ions by the first ion optics inside the  MBMS. This is resolved by keeping the orifice diameter as small as possible while allowing for acceptable measurement times. This balance between Debye length and orifice diameter is especially difficult for streamer plasmas, where the very high electron density at the streamer head allows the penetration of these ionisation waves dynamically, even into the very small orifices. Nevertheless, it may be possible to deduce any plasma penetration from the measured data. When the energy is determined by the seeding in the supersonic expanding neutral gas stream, any plasma penetration into the MBMS is not likely. Since the temporal resolution of the measured IEDFs is on the microsecond scale at most for a given signal-to-noise ratio, the occurrence of a burst of high energetic ions when a streamer impacts the surface on a nanosecond time scale may even be hidden in the microsecond temporal averaging of the data. 

Even in the case of a perfect MBMS sampling instrument, the temporal and energetic structure of the IDEF from an impacting streamer may even be very complex. At first, a streamer impacts the surface, and any energetic ions may contribute to the incident species flux. However, after the plasma impacts, a quasineutral plasma channel remains, and a positive space charge region is found right in front of the surface. After the streamer impact, ion-electron recombination sets in with a time constant of less than 100 ns at atmospheric pressure. This recombination is not possible directly in front of the surface since the electrons are lost to the surface. Then, the space charge region decays by drift and diffusion of the residual ions. As a consequence, a plasma glow and a space charge region reside in front of the surface for a time span of several microseconds. This has been seen in various experiments on streamer plasmas interacting with the surface or in plasma actuators, wherever a streamer plasma propagates next to a surface. Consequently, a long-lasting ion flux to the surface is expected, being a fingerprint of a decaying space charge region, rather than the direct streamer plasma impact. 

Summarizing, one may state that the nature of the IEDF can be dominated by various aspects:  (i) high energy ions from a streamer, (ii) high energy ions due to plasma penetration, (iii) low energy ions, due to collisional sheaths, (iv) shift in the energy due to surface charging, (v) long lasting ion fluxes due to a decaying space charge region in front of the surface.

To analyse the IEDFs, we use molecular beam mass spectrometry (MBMS) with an energy filter to study the ions from a dielectric barrier discharge plasma jet. The plasma setup follows a previously published design \cite{chauvet_characterization_2014}. The species are sampled from the plasma with a 40 $\mu$m metallic orifice or with a 50 $\mu$m ceramic orifice. Similar experiments have been performed with the very same setup by Große-Kreul \cite{grosse-kreul_mass_2015} analysing the effluent of an RF plasma jet, where residual water cluster ions, in particular, were being monitored, but no direct plasma impact on the surface occurred. With these experiments, we address the question of which energies are observed in the ion energy distribution, and whether energy-resolved, MBMS can adequately monitor the temporal evolution of plasma chemistry.  
 
\newpage

\section{Experiment}

\subsection{Plasma reactor and plasma performance}

\begin{figure}
    \centering
\includegraphics[width=0.6\textwidth]{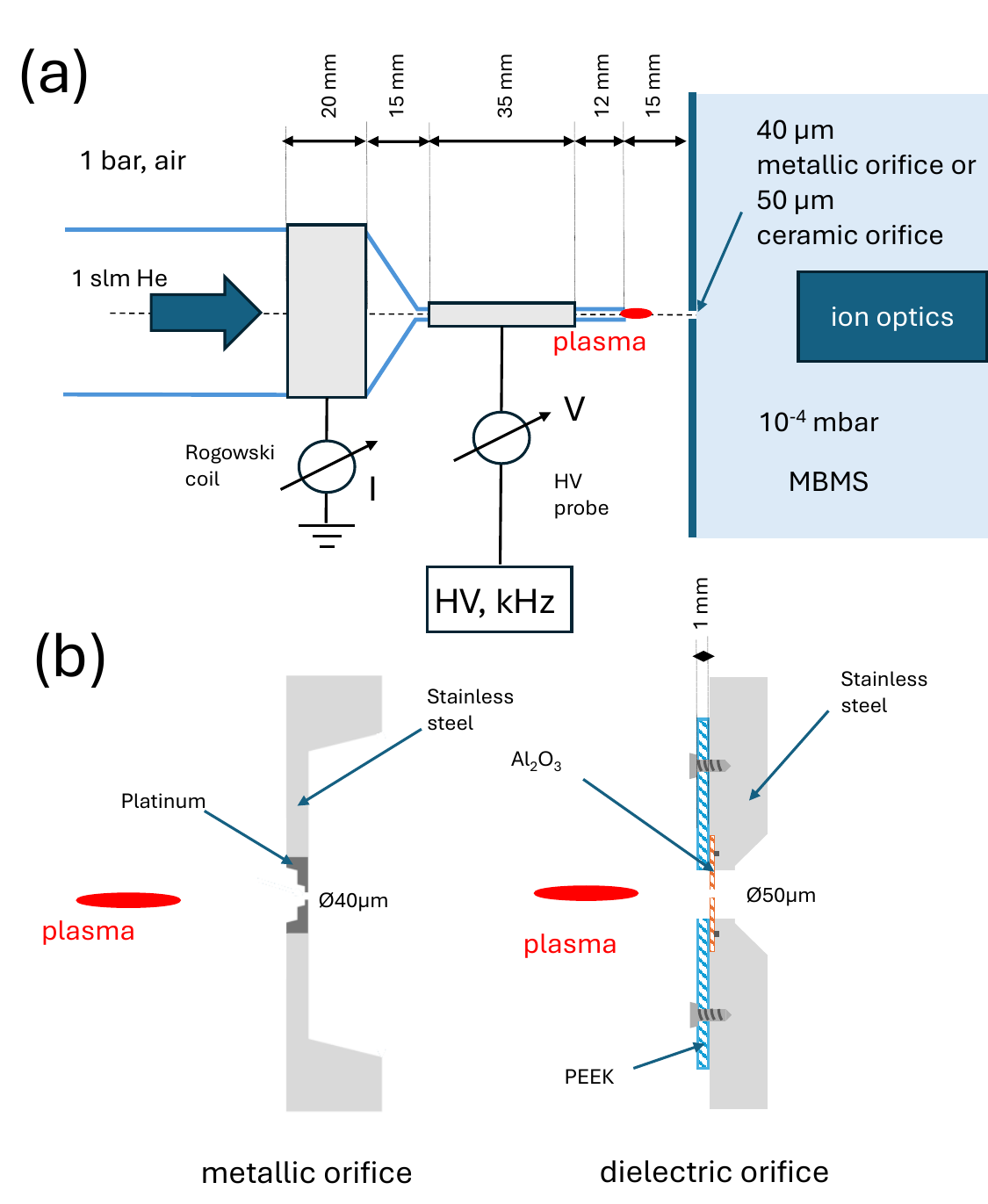}
    \caption{(a) Schematic of the experimental; (b) Close-up of the metallic (left) and ceramic (right) orifice; }
    \label{fig:setup}
\end{figure}

A schematic of the plasma setup is shown in Fig. \ref{fig:setup}a. The plasma source consists of an asymmetric quartz reactor, composed of a capillary with an outer diameter of 4 mm and an inner diameter of 2 mm, which widens at the back end to a 38 mm inner diameter cylinder. A powered electrode, 35 mm in length, is wrapped around the capillary, while a grounded electrode encircles the larger cylindrical part at an axial distance of 15 mm from the powered electrode. The glass capillary extends beyond the powered electrode by 12 mm and is positioned 15 mm apart from the orifice of the mass spectrometer. A gas flow of 1 slm Helium is used, yielding a flow velocity inside the capillary of 5.3 m/s. The system is operated with a kV, kHz power supply, producing high-voltage bursts with a repetition frequency of 20 kHz. 

The voltage and current are measured with a voltage probe in the line to the powered electrode and with a current probe in the line to the grounded counter electrode. The capacitive loading of the system yields a displacement current, which has to be subtracted from the total current to yield the actual dissipated current. For this, the VI signal $U_{\rm meas}$ and $I_{\rm meas}$ are being fitted with:

\begin{figure}
    \centering
\includegraphics[width=0.6\textwidth]{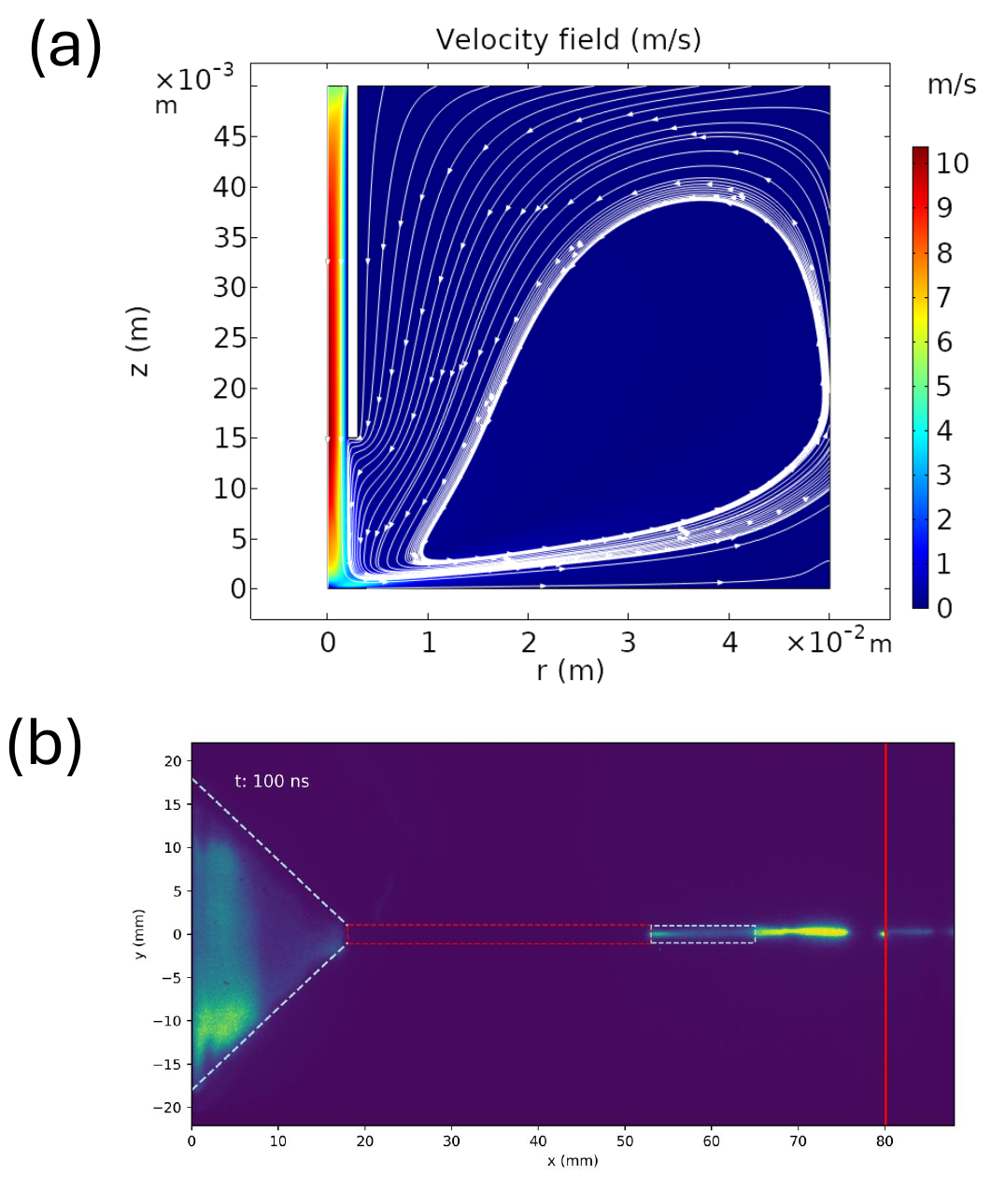}
    \caption{(a) Fluid simulation of the velocity streamlines, the colour scale shows the flow velocities; (b) ICCD images, the blue dashed lines indicate the glass capillary, the red dashed line the powered electrode, the red vertical line the position of the orifice.}
    \label{fig:setupw}
\end{figure}

\begin{equation}
\langle P \rangle = f \int_{T}U_{\rm meas}\left(I_{\rm meas}-C_{\rm sys}\frac{dU_{\rm meas}}{d t}\right)dt
\label{eq:power}
\end{equation}

The fit yields a system capacitance of $C_{\rm sys}$ = 5.1 pF. The average plasma power can also directly be deduced by integrating over the product of $U_{\rm meas} \times I_{\rm meas}$ since the capacitive part automatically cancels. However, this fit is required to determine the time dependence of the absorbed power. The system operates in an ambient atmosphere, so nitrogen and oxygen are entrained in the flow at the exit of the helium gas jet. Fig. \ref{fig:setup}a shows a simple fluid simulation of the flow pattern using COMSOL, where the helium flow is modelled as it enters the capillary at a velocity of 5.3 m/s corresponding to a flow of 1 slm and exits the capillary on the central axis before impacting the surface at a distance of 15 mm. A prominent vortex is formed in front of the surface, which efficiently entrains the ambient air into the helium gas jet.

The plasma evolution is imaged using an ICCD camera (PI-Max 3), with an exposure time of 10 ns and triggered by the HV probe to be in sync with the plasma generation. 1000 images are accumulated, and a typical image is shown in Fig. \ref{fig:setup}b at $\Delta t$=100 ns after the maximum of the positive voltage pulse: The conical shape of the quartz cylinder is visible at the back end of the capillary, the red dashed square marks the powered electrode, and the blue square the capillary. The red line marks the location of the surface. One can see the bright emission beyond the exit of the capillary.  

\subsection{Ion sampling}

\subsubsection{Design of the MBMS}

% -Several sampling related effects altering the measurements 
 %-distinguishing sampling related effects from plasma related effects

A molecular beam mass spectrometer (MBMS) is used to sample ions impinging on the surface. The MBMS system has been developed by Große-Kreul \cite{grosse-kreul_mass_2015} and is described in detail there. Here, we only briefly describe its elements. Ions enter a first pumping stage ($10^{-4}$ mbar) through a sampling orifice made from either a metal or a dielectric. A schematic of the orifice designs can be seen in Fig. \ref{fig:setup}b. The metallic orifice consists of a conical platinum aperture 40 $\mu$m in diameter that is pressed onto a stainless steel plate. The dielectric orifice consists of a 0.13 mm thick plate made of alumina ceramic with an orifice 50 $\mu$m in diameter. It is clamped on a stainless steel plate between a PEEK plate of 1 mm thickness using an O-ring for vacuum sealing and to avoid damaging the ceramic by mechanical stress. When species from the ambient pass the orifice in the first pumping stage, they undergo supersonic expansion and a molecular beam is formed. 

\begin{figure}
    \centering
\includegraphics[width=0.6\textwidth]{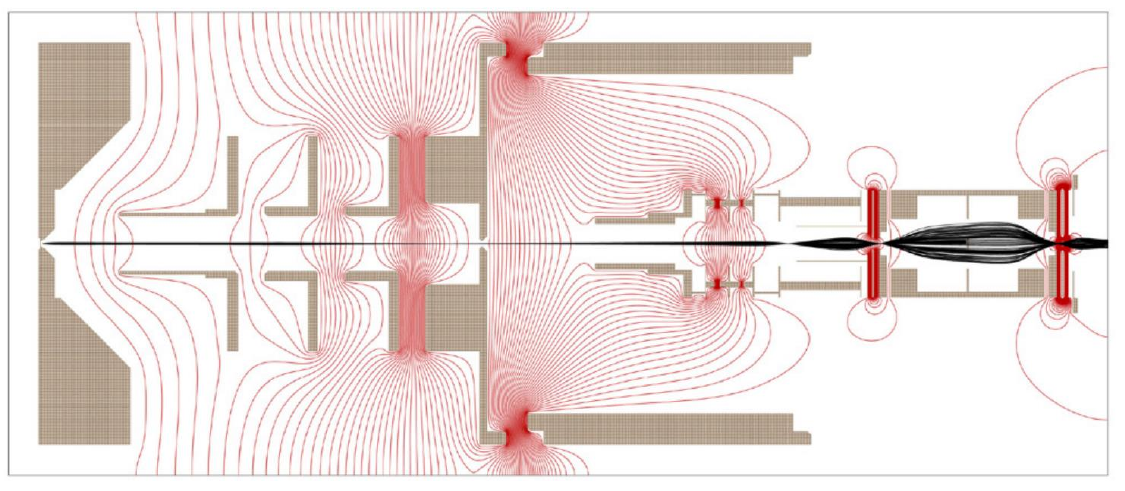}
    \caption{Simion simulation domain, showing the ion trajectories inside the first ion optics, followed by the sampling optics of the HIDEN PMS spectrometer. With permission from \cite{grosse-kreul_mass_2015}.}
    \label{fig:ionoptics}
\end{figure}

A set of four electrostatic lenses is employed in the first pumping stage to prefocus the ions to reduce any signal loss (see Fig. \ref{fig:ionoptics}). The lens voltages are tuned to maximize the ion transmission using oxygen ions as a reference species. Consequently, ions with a largely different mass than oxygen are detected with a smaller sensitivity. However, the shape of the energy distribution functions of all measured ions remains unaffected. In the second pumping stage, a Hiden PMS ion spectrometer is mounted with a Bessel box energy analyser used to select specific ion energies, followed by a quadrupole for mass analysis before the ions are detected on a secondary electron multiplier (SEM). The energy is scanned by varying a reference voltage $U_{\rm ene}$ at the entrance of the Bessel box, which is also the reference voltage for the remainder of the ion trajectory through the quadrupole. The ion trajectories had been simulated using SimIon, as illustrated by the electric fields in Fig. \ref{fig:ionoptics} from \cite{grosse-kreul_mass_2015}. In general, the energy scale must be calibrated to connect the specific setting of $U_{\rm ene}$ to the case of $E_{\rm kin}$ = 0 eV. The value for $U_{\rm ene}$ = $E_0$ defining the setting for $E_{\rm kin}$ = 0 eV is determined from the analysis of water cluster ions, as shown below. Time-resolved measurements were performed in sync to the pulsed plasma excitation by using a multi-channel scalar card to distribute the ion count rates on a time axis. 
%\textcolor{red}{TODO: MS inlet + Beam schematic}

%\begin{table}[!ht]
%    \centering
%    \caption{Ion optics voltages. $U_{1-4}$ respond to the voltages applied to the external ion lenses.}
%    \begin{tabular}{|l|l|}
%    \hline
%        U$_{ene}$ & x / V \\ \hline\hline
%        U$_{1}$ & -10  \\ \hline
%        U$_{2}$ & -12  \\ \hline
%        U$_{3}$ & -35  \\ \hline
%        U$_{4}$ & -250  \\ \hline
%        U$_{ext}$ & -11  \\ \hline
%        U$_{lens}$ & -95  \\ \hline
%        U$_{cage}$ & x  \\ \hline
%        U$_{source}$ & -17 +x \\ \hline
%        U$_{end}$ & -12.5 + x \\ \hline
%        U$_{cyl}$ & -1.4 + x \\ \hline
%        U$_{foc}$ & -20 + x  \\ \hline
%        
%    \end{tabular}
%    \label{tab:IonOptics}
% \end{table}

%-external ion optics
%-mass dependent tuning (inertia), taking m=32, as most ambient species are expected in this range
%-brief description of chromatic aberration?? ( more important for large energy differences of 10s of eV) so maybe not...

%TODO

%Design of the orifices metallic = 40 µm, ceramic = 50 µm
%-brief description

\subsubsection{Sampling from a supersonic expanding beam}

The species are sampled via a metallic ($d_{\rm orifice}$ = 40 $\mu$m ) or ceramic ($d_{\rm orifice}$ = 50 $\mu$m) orifice and are accelerated into the low-pressure first stage. Due to collisions in the beam, all species are accelerated to the very same velocity, determined by the thermodynamics of the expanding gas. The maximum mean beam velocity can be approximated as \cite{Morse_1996}:  

\begin{equation}
v_{\rm beam} = \sqrt{\frac{2\gamma}{\gamma-1}\frac{k_BT}{m}}
\end{equation}

with $\gamma$ the adiabatic coefficient, $T$ the gas temperature and $m$ the mass of the expanding species. If we take the values for helium, we obtain a beam velocity of $v_{\rm beam}$ = 1800 m/s. Due to collisions with the carrier gas, all species will reach the same final velocity. Consequently, their energy $E_{\rm beam}$ scales simply with the mass $m$ of the species by a scaling factor $c$ since their velocity is identical:

\begin{equation}
E_{\rm beam} = \frac{1}{2}m v_{\rm beam}^2 = c m\label{eq:ebeam}
\end{equation}

The analysis of the energy spectra of seeded molecular ions can be used to identify a scaling of $E_{\rm beam} = c m$. This constitutes a test of proper beam seeding. Any chemical reaction among the species during supersonic expansion does not affect this scaling because, independent of the species' mass, their final velocity remains $v_{\rm beam}$. The position of the \textit{virtual source} downstream, the quitting surface, where the flow transitions into the molecular regime, can be calculated by \cite{huang_thermodynamic_2018}:

\begin{equation}
z_{\rm quitting} = -d_{\rm orifice}\ln\left(-\frac{mv_{\rm beam}^2}{2k_BT\left(1+\frac{1}{\gamma-1}\right)}\right)
\end{equation}

With $d_{\rm orifice}$ = 40 $\mu$m, $v_{\rm beam}$ = 1580 m/s, $\gamma=5/3$ for helium, and $T$ = 300 K, we obtain $z_{\rm quitting}$ = 60 $\mu$m. Beyond the quitting surface, collisions are still possible, but the collision rate decreases exponentially over two orders of magnitude over a distance of 10 times the orifice diameter \cite{Morse_1996}. The beam will eventually interact with the background gas until a shock front is formed, the so-called Mach disk. Its position can be calculated as:

\begin{equation}
z_{\rm Mach} = 0.67 d_{\rm orifice} \sqrt{\frac{p_0}{p_b}}
\end{equation}

This yields a distance of $z_{\rm Mach}$ = 85 cm for a background pressure $p_b$ = 10$^{-4}$ mbar and an initial beam pressure of $p_0$ = 10$^5$ Pa. This position lies in the second pumping stage in a region of much lower pressure of 10$^{-8}$ mbar so that any shock front formation is avoided and a freely expanding molecular beam is assured.

%\subsubsection{Calibration of the energy and time scale of the MBMS}
%The energy scale of the MBMS needs to be calibrated to correlate ion energies with the voltage applied to the energy filter ($U_{\rm ene}$). Water cluster ions are used for this purpose, as they appear in abundance over a large range in mass.
%unsure about this section, should refer to the process as done by Simon. But as this is kind of a big point later, discussion here might be redundant
%To analyze the energy scale and the 

%The actual measurement of the ion energies in the HIDEN PMS is 

%\begin{figure}
%    \centering
%\includegraphics[width=1\textwidth]{BulletJetFigures 3.pdf}
%    \caption{SIMION simulation domain (a). Voltages applied to the individual lenses and elements, the coloured lines describe the variation with voltage $U_{\rm ene}$ (b).}
%    \label{fig:simiondomain}
%\end{figure}

%$U_{\rm ene}$ as scale, problem of finding the zero on the energy scale etc.equation

\newpage
\clearpage

\section{Results}

\subsection{Plasma dynamic}

\begin{figure}
    \centering
\includegraphics[width=0.6\textwidth]{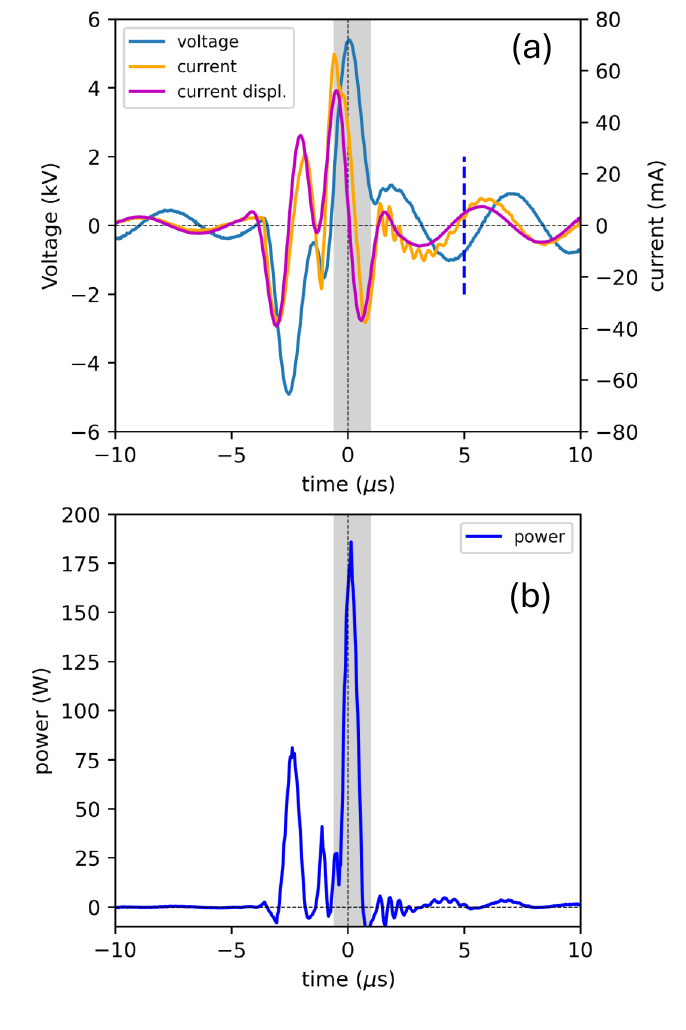}
    \caption{(a) Voltage (blue) and current (orange) of the dielectric barrier discharge plasma jet. The displacement current (magenta) is deduced from the analysis of the VI characteristics after 5$ \mu$s (dashed blue line). The grey area indicates the period when the ICCD images are taken; (b) dissipated power.}
    \label{fig:vi}
\end{figure}

At first, we regard the plasma dynamics as monitored by the VI probe and the synchronized ICCD images. Fig. \ref{fig:vi}a shows the voltage (blue line) and current (orange) measurements for a plasma using a 10 kV peak-to-peak voltage pulse. The grey area shows the time window in which the ICCD images are taken. The signal is fitted with a sinusoidal at times later than 5 $\mu$s (indicated by a dashed blue line in Fig. \ref{fig:vi}) to yield the displacement current (magenta line). The absorbed power is calculated from this, as shown in Fig. \ref{fig:vi}b. One can see two phases of absorbed power, one for the negative voltage peak and one for the positive voltage peak.

\begin{figure}
    \centering
\includegraphics[width=0.6\textwidth]{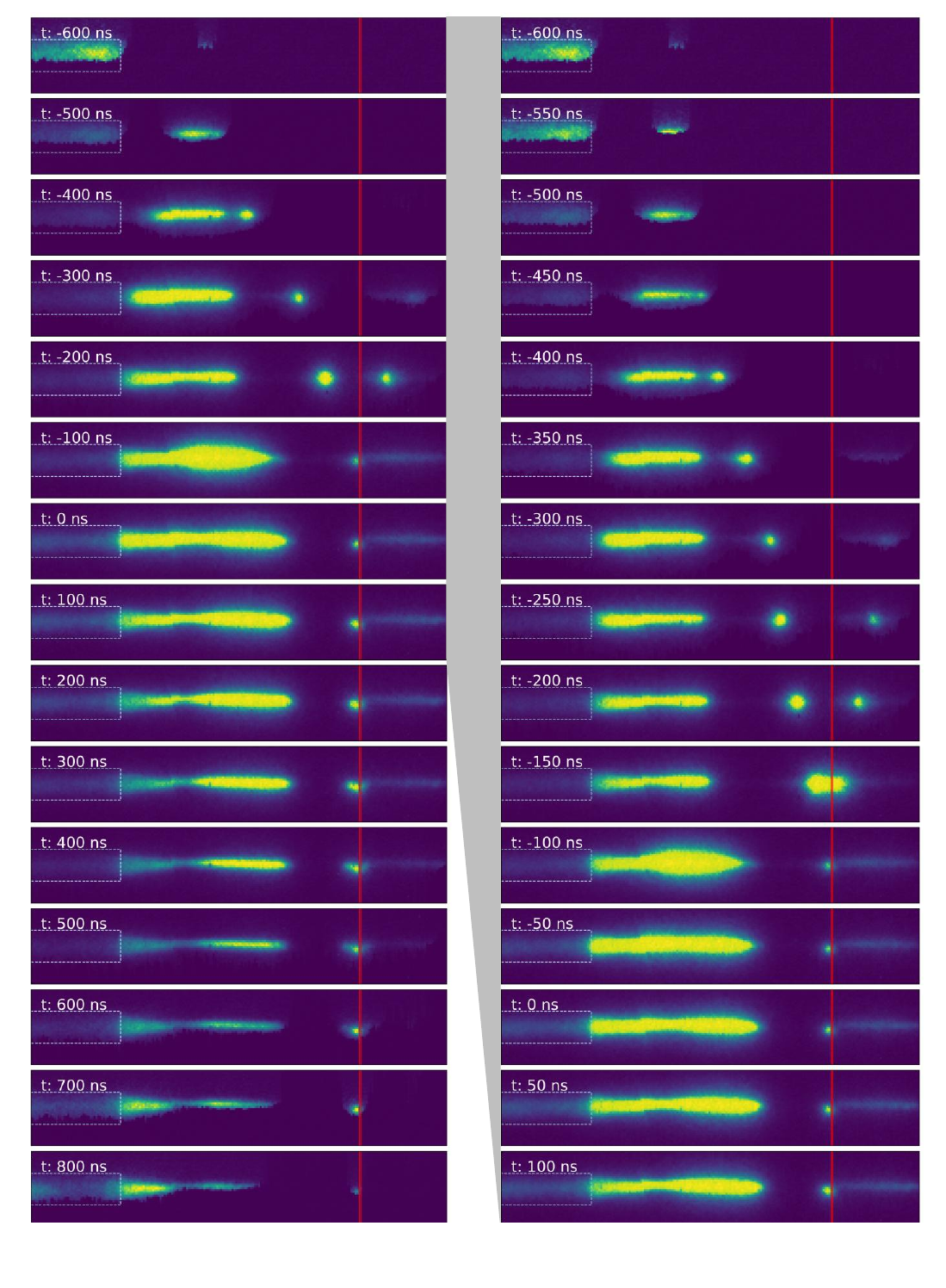}
    \caption{ICCD emission at different times during the positive voltage pulse. A blue dashed line indicates the glass capillary, and a red line indicates the orifice. The times are indicated, with $t$= 0 s corresponding to the maximum of the positive voltage pulse. The left panel shows the period between $t$=-600 ns to $t$=800 ns, and the right panel shows the bullet propagation phase more closely between $t$=-600 ns to $t$=100 ns (connected by a grey area).}
    \label{fig:iccd}
\end{figure}

The ICCD camera images revealed that the plasma emission is maximal between the powered and grounded electrodes during the negative voltage peak, but maximal between the powered electrode and the orifice in the positive voltage peak. The zero of the time axis is set to the maximum of the positive voltage peak, defined as $t=0$ s. The temporal evolution of the ICCD images starting at $t=-0.6$ $\mu$s at the beginning of the positive voltage pulse is shown in Fig. \ref{fig:iccd} (a grey area in Fig. \ref{fig:vi}). The left panel shows the time span between $t$=-600 ns to $t$=800 ns, and the right panel shows the bullet propagation phase more closely between $t$=-600 ns to $t$=100 ns. A bright plasma extends downstream and \textit{outside} beyond the capillary tip. The excited helium plasma species interact with the entrained nitrogen and oxygen at this location. Since excited nitrogen and oxygen species provide many more emission lines than helium, the ICCD images become intense in the first 5 to 10 mm beyond the end of the capillary. One could also note that a small \textit{plasma bullet}, an ionization wave, forms at $t$ = -450 ns that swiftly propagates towards the orifice until $t$ = -150 ns. From the dimension of the setup and the time steps of the ICCD images, one can deduce the velocity of this ionization wave of 10$^4$ m/s, which is consistent with the literature\cite{Lu_2014}, when the plasma bullet approaches the orifice at $t$= -200 ns, its intensity increases (one could also notice a bright spot behind the orifice, which is just a reflection from the metallic front of the MBMS). After the plasma bullet impacts the surface at $t$=-150 ns, a steady emission is visible until $t$ = 800 ns. The same emission is observed with the bullet impinging on a solid metallic plate (see supplementary material) and is thus not related to plasma penetration into the orifice but rather the decay of a space charge region left after the bullet's initial impact. The plasma lasts for 1.4 $\mu$s, equivalent to the period of the positive voltage pulse. Plasma emission is always related to the presence of energetic ions. Therefore, even after plasma emission ceases, residual electrons and ions may still exist until they recombine with each other or with adjacent surfaces. These recombination processes have been shown to last several 100 ns \cite{mohsenimehr_plasma_2025}.
%\textcolor{red}{TODO: refer to supplementary material and Liu paper for the remaining emission liu interaction, ion wind argument secondary emission ?}
In separate experiments (not shown), the distance between the capillary exit and the MBMS has also been reduced to 1 cm. In this case, the erosion of the orifice was so significant that its diameter increased significantly over time. At a distance of 1.5 cm, the plasma channel - located between the nozzle and the bullet - does not reach the MBMS. Only the plasma bullet impacts the orifice, delivering a much lower thermal load and allowing for long measurement times without affecting the performance of the MBMS system. 

%\textcolor{red}{does it makes sense to analyse the absolute signal intensities}

\subsection{Ions from the DBD plasma jet}

\subsubsection{Overview ion spectrum}

\begin{figure}
    \centering
\includegraphics[width=1\textwidth]{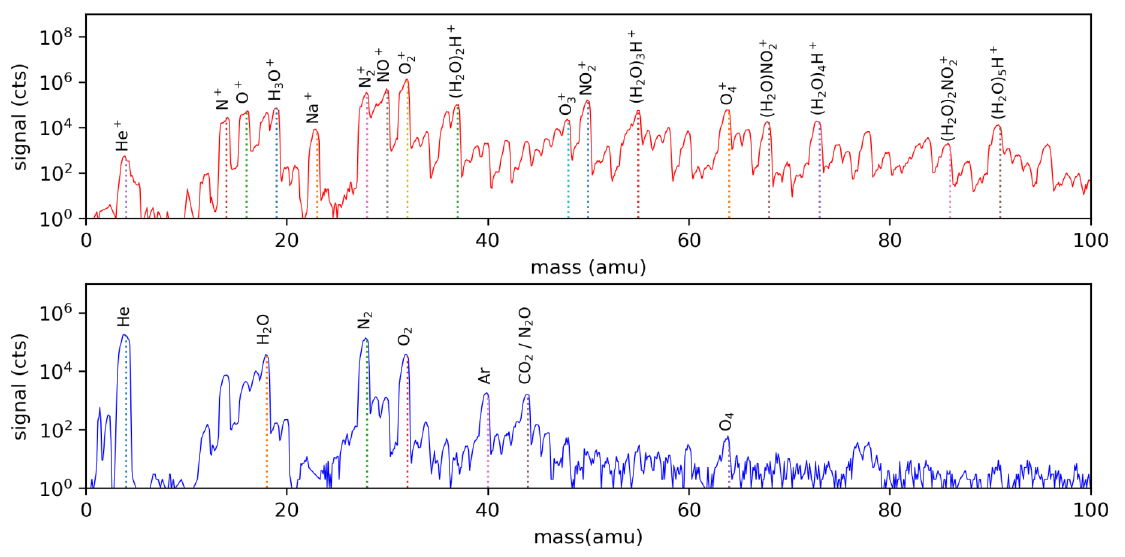}
    \caption{Overview spectrum of positive ions (a) and neutrals (b). The individual peaks are assigned. Measured at $U_{ene}$ = -3.4 V.}
    \label{fig:ionoverview}
\end{figure}

Fig. \ref{fig:ionoverview} shows the overview spectra of positive ions and neutrals (with the electron energy in the ioniser of the MBMS set to 70 eV) measured for a plasma jet at 10 kV peak-to-peak voltage. The dominant ions are N$_2^+$, O$_2^+$, NO$^+$ and the series of water cluster ions (H$_2$O)$_n$H$^+$. A small peak at mass 23 might indicate the presence of Sodium ions, which may originate as a contaminant from the glass capillaries being used in the experiment (any Si mass peaks are hidden underneath the N$_x$O$_y^+$ peaks). The ion spectrum extends to much higher masses than the neutral spectrum. This composition is very typical for such plasmas and has also been measured by others \cite{jiang_absolute_2021,ito_mass_2015,oh_investigating_2015}. However, a notable feature is the abundance of He$^+$ ions observed in the spectrum. At atmospheric pressure, these ions only have a short lifetime due to collisions with other particles. The more commonly measured He$_2^+$ created in such collisions is absent in the measured spectrum. Such an observation could indicate that He$^+$ ions are created on the low-pressure side due to the penetration of the streamer through the orifice. In that case, however, one would expect to observe ions at energies significantly higher than the energy dictated by supersonic expansion. The dominance of He$_2^+$ ions holds for pure helium plasmas at atmospheric pressure. In our experiment, however, significant charge exchange to molecular nitrogen and oxygen occurs, so the sampled ion composition may reflect the different efficiencies for charge exchange. For example, if charge exchange from He$_2^+$ is much more dominant than from He$^+$, the ion composition could easily be explained. Any detailed data on these charge exchange cross-sections is missing. Finally, the He$^+$ signal is only of the order of 10$^{-3}$ of the dominant NO$_x^+$ ions, so any deeper reasoning about differences between He$^+$ and He$_2^+$ densities is beyond the scope of the paper.
%\textcolor{red}{TODO Daniel: read Yuan Paper computational study... for absense of He2+; Cross sections Charge transfer in lee$_197$; Nevertheless, the energies follow supersonic expansion well, so plasma from inside the MS doesnt seem likely....}

\subsubsection{Water cluster ions}

\paragraph{Energy spectrum}~\\

\begin{figure}
    \centering
\includegraphics[width=0.5\textwidth]{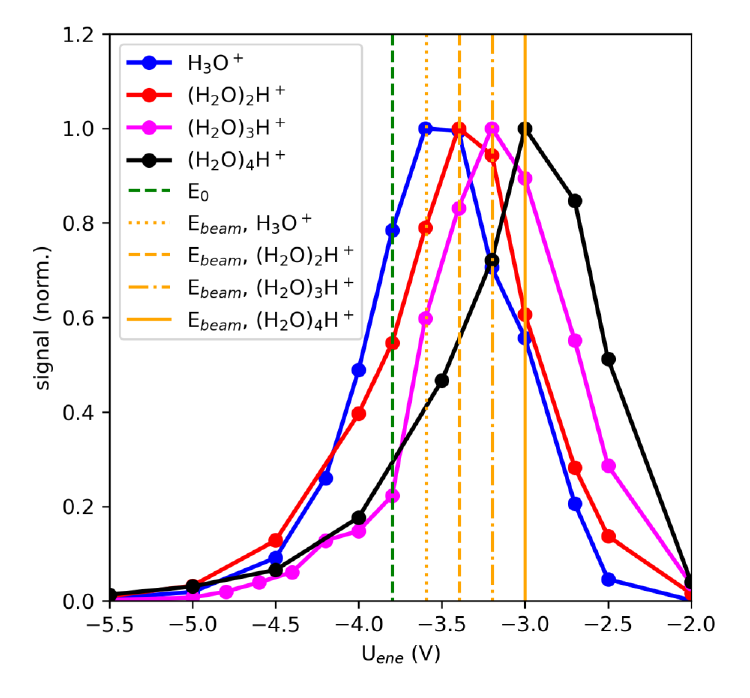}
    \caption{Energy distribution of dominant water cluster ions. The green vertical line corresponds to $E_0$, and the orange lines correspond to the prediction of water cluster ion energies according to eq. \ref{eq:escaling}. }
    \label{fig:energywatercluster}
\end{figure}

To assess the energy scaling of the ion sampling, we first regard water cluster ions. Fig. \ref{fig:energywatercluster} shows the energy distribution of the different water cluster ions \waterone, \watertwo, \waterthree, \waterfour as a function of the voltage setting for $U_{\rm ene}$. It is striking that the maxima are equally separated by $\Delta E$  = 0.2 eV. In the simplest estimate, we assume that these ions start at the orifice of the mass spectrometer with a thermal kinetic energy $E_{\rm kin}$ $<$ 1 eV. These water cluster ions are then accelerated in the supersonic expansion after passing the orifice. Due to collision with Helium, all these species acquire the same velocity $v_{\rm beam}$ and energy $E_{\rm beam}$ according to eq. \ref{eq:ebeam}. Due to these collisions, any information on their initial energy is lost. This simple estimate can be tested by regarding the energy scaling of the peaks plotted against $U_{\rm ene}$, which follows a relation as:

\begin{equation}
U_{\rm ene} [{\rm eV}] = E_0 + E_{\rm beam} = E_0 + c m_{\rm  (H_2O)_nH^+} \label{eq:escaling}
\end{equation}

The values $E_0$ = -3.8 eV and $c$ = 0.011 eV/amu yield good agreement for the peak positions of different water cluster sizes $\rm (H_2O)_nH^+$ with $n$=1..4. The fitted value for $c$ is consistent with the previous calibrations by Große-Kreul of $c$ = 0.012 eV/amu \cite{grosse-kreul_mass_2015}. If we assume helium as the main carrier gas, we obtain a velocity $v_{\rm beam}$ of 1800 m$^{-1}$, but the scaling of the water clusters expressed as $c$ can be converted to a beam velocity of only 1580 ms$^{-1}$. This is not in contradiction because the helium beam expands in air, and a fraction of air entrainment is expected. Assuming a mixture of 96\% helium and 4\% nitrogen, we obtain a slightly higher mass of the expanding gas, leading to a smaller beam velocity of $v_{\rm beam}$ = 1580 m/s. An energy of $U_{\rm ene}$ = -3.8 eV should correspond to ions at zero kinetic energy collected by the extraction ion optics before injection into the energy filter. This value corresponds to the zero value $E_0$ of our energy scale. The data are in agreement with the scaling, expressed by eq. \ref{eq:escaling}, which suggests that the sampling follows the theory of seeded molecular beams and that the detected energy of the species originates \textit{only} from the acceleration during supersonic expansion. 

%The zero of the energy scale $U_{\rm ene}$ is at $E_0$ = -3.8 eV is also consistent with the energy spectrum of the neutrals, which do not show the drift in energy since all species are starting in the ionizer at rest. %\textcolor{red}{new orifice shift by 0.4 eV}. %The same energy for different mass neutrals also indicates that the beam component of neutrals from sampling is apparently quenched, and the signal is dominated by the thermalized background gas density in the ionizer.

%\textcolor{red}{DECIDE WHETHER WE USE SIMION at ALL}

%\textcolor{red}{The energy scale and its shifts are analyzed by a SIMION simulation, which uses ions starting at the quitting surface with an opening angle of 6° and an initial energy corresponding to the energy from the supersonic expansion. The resulting trajectories are followed, and the resulting signal is plotted in Fig. \ref{fig:simtransmission}. The transmission of the quadrupole is not taken into account.} %The transmission reproduces peaks at equivalent distances as in the experiment. The simulation also yields a low-energy flank.

%\begin{figure}
%    \centering
%\includegraphics[width=0.5\textwidth]{BulletJetFigures 8.pdf}
%    \caption{Simulation of the transmission function of different water clusters. \textcolor{red}{why is the scaling of the 2D energy vs. time maps different}}
%    \label{fig:simtransmission}
%\end{figure}

%\newpage
%\clearpage

\paragraph{Time sequence of the impacting species ions}~\\

\begin{figure}
    \centering
\includegraphics[width=0.8\textwidth]{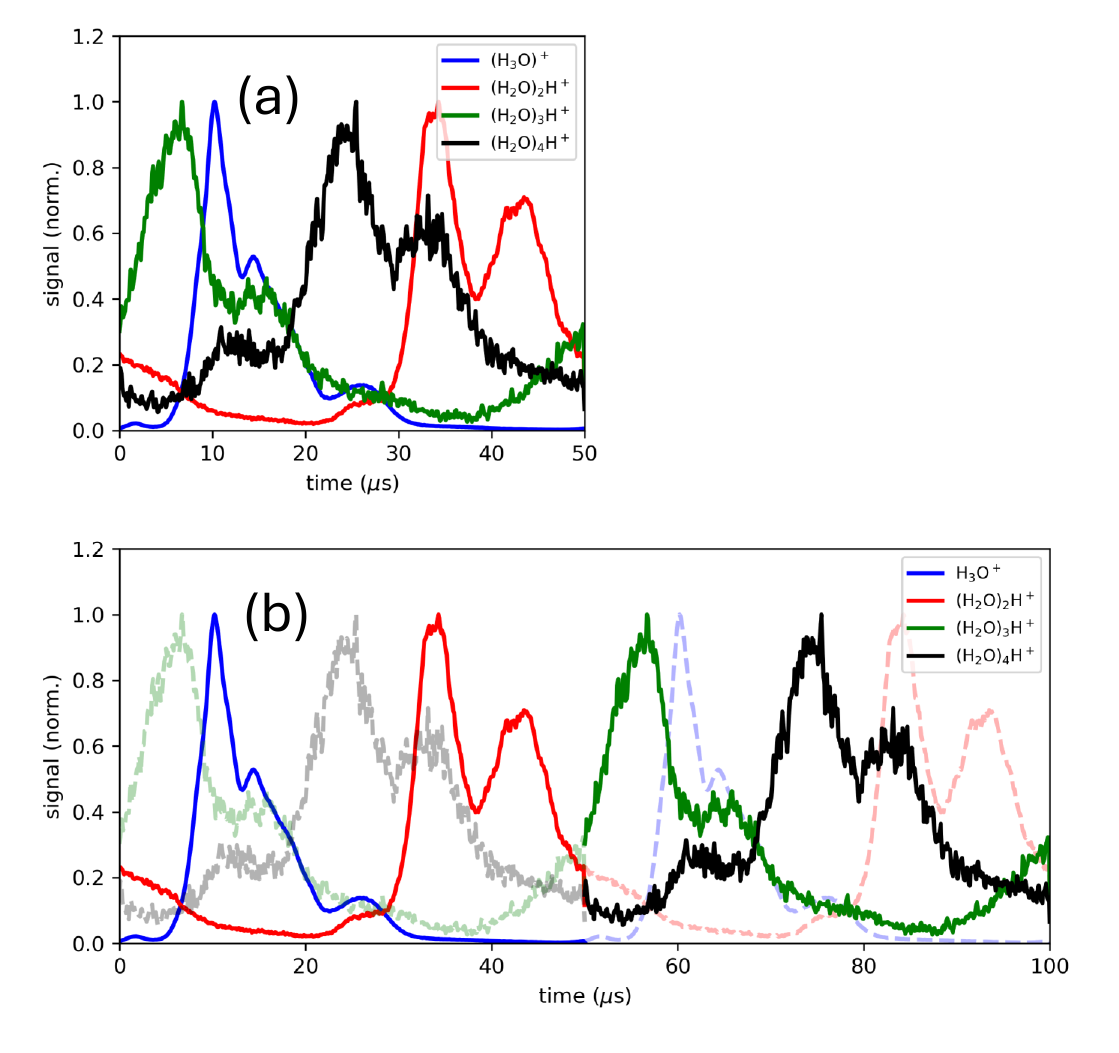}
    \caption{(a) signal of the different water cluster ions arriving at the detector sampled at their energy $E_{\rm beam}$; (b) the signals for \waterthree and \waterfour are shifted by 50 $\mu$s to later times. The transparent signals indicate the signal in the earlier or later 50 $\mu$s time window of the synchronized sampling.}
    \label{fig:watercluster}
\end{figure}

Next, we regard the time dependence of the signals for the different water clusters measured at their individual energies $E_{\rm beam}$. This is shown in Fig. \ref{fig:watercluster}a. One can see that the time dependence stretches the complete cycle of the high-voltage pulses, and several peaks can be seen.  

Given that the impact of the plasma bullet occurs within a few nanoseconds and that the plasma afterglow lasts around 1.5 $\mu$s, the majority of ions are expected to enter the sampling system at the same time with a delay of a few $\mu$s at most. However, they are detected at vastly different times at the SEM, and the signals of the different water clusters appear to be shifted randomly in time to each other (Fig. \ref{fig:watercluster}a). However, one must remember that the ions are generated in plasma pulses every 50 $\mu$s and the trigger of the multi-channel scalar card to register the data as well, so the time coordinate should be interpreted modulo 50 $\mu$s. By correcting the time scale using modulo 50 $\mu$s for each of the water clusters, we get a rather regular sequence of water clusters where the lightest clusters appear the earliest in the MBMS, followed by the heavier ones (Fig. \ref{fig:watercluster}b, the data for \waterthree and \waterfour are shifted by 50 $\mu$s in time). This correction places the detection times of the different water clusters by the MBMS on an absolute time scale. 

At first, one might argue that the different detection times of the water clusters at the SEM detector in the mass spectrometer are surprising, given that they are expected to start all simultaneously when the plasma bullet hits the orifice. However, this can be resolved by considering the different time of flights (TOF) for species with different energies and masses in the detection system, as analysed in the following. The TOF is accumulated in the energy filter and the mass filter. The mass dependence mainly causes a variation of the TOF during the transfer of the ions in the mass filter at a small kinetic energy by design. For example, a typical transfer energy in the quadrupole is $E_{\rm transfer}$ = 3 eV (in the typical range for QMS from 1 to 10 eV). If we assume a length $d$ of the quadrupole to the detector, we obtain a TOF scaling in the QMS of $t_{\rm TOF, QMS} = d \sqrt{m/(2 E_{\rm transfer})}$. The TOF, up to the entrance of the QMS, through the ion optics and the energy filter, has to be added. Since the kinetic energies in the energy filter are much higher due to the acceleration of the ions, much shorter TOFs are expected. As a simple estimate, we use a scaling of the TOF according to: 

\begin{equation}\label{eq:tofscaling}
\Delta t_{\rm TOF} = \Delta t_{\rm TOF,0} + \sqrt{\frac{m}{2E_{\rm transfer}}}d
\end{equation}

\begin{figure}
    \centering
\includegraphics[width=0.5\textwidth]{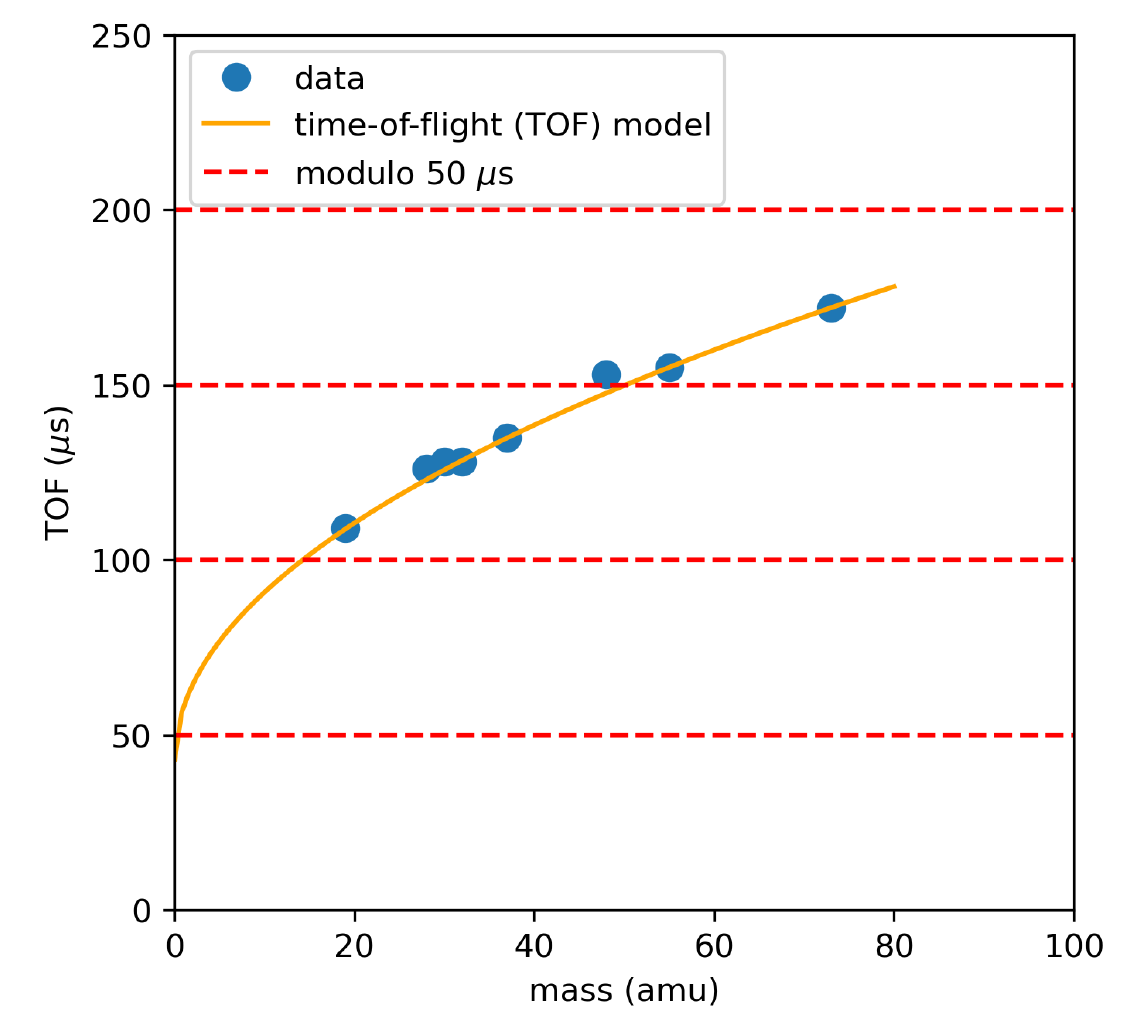}
    \caption{Time of flight of the different water clusters (blue points) in the MBMS system versus ion mass. The solid line denotes a fit according to eq. \ref{eq:tofscaling}. The red dashed lines denote the modulo 50 $\mu$s time interval of the triggered signal acquisition.}
    \label{fig:tofsim}
\end{figure}

If we use $E_{\rm transfer}$ = 3 eV, $d$ = 0.37 m, and $\Delta t_{\rm TOF,0}$ = 43 $\mu$s, we obtain an excellent fit of the arrival times of the different water clusters at the detector, as shown in Fig. \ref{fig:tofsim}. For the sake of the fit, a 100 $\mu$s delay (according to the modulo 50 $\mu$s) has been added to the time at which a signal of the respective species was measured. This scaling is only a simple estimate, and a more thorough analysis using ion optics software such as SIMION is possible and has been performed in the past \cite{GrosseKreul_phd}. Here, we use the mass scaling of the different water clusters directly, covering the mass range of the oxygen- and nitrogen-containing ions. 

%. Here, the mass dependence is much smaller since much higher voltages are used in the ion optics. The TOF $t_{TOF, energy filter}$ = 43 $\mu$s. This gives the total TOF $t_{\rm TOF} = t_{TOF, energy filter} + t_{TOF, QMS}$ as shown by the blue line in Fig. \ref{fig:tofsim}. Given the temporal position of the prominent peaks for the different water cluster ions being modulo 50 $\mu$s, a very good agreement is obtained.

%One can see that the TOF vary strongly with the energy and mass of the sampled ions. The lines indicate the TOF for ions that start simultaneously but have different energies at the quitting surface according to \ref{eq:escaling} and different mass. In the energy range of $U_{ene}$ >-1.2 eV, the TOF scaling yields a time shift of 250 $\mu$s for an energy difference of 0.4 eV. In the energy range energies below $U_{ene}$<-1.2 eV, the TOF scaling is 0.45 eV, corresponding to 10 $\mu$s. Since most of the data are acquired in the latter $U_{\rm ene}$ energy range, The scaling of 0.45 eV versus 10 $\mu$s is relevant for interpreting the data in the following.

The uniqueness of the fitted values for $E_0$ and $c$ from the time-integrated energy spectra is limited because the energy distributions' resolution is only 0.2 eV. The time-dependent data presented in the following reveal that each energy distribution comprises several peaks measured at different times. The consistent explanation of all these peaks requires values of $E_0$ = -3.9 eV and $c$ = 0.012 eV/amu, as discussed below.

\paragraph{Energy vs. time maps of the impacting water cluster ions}~\\

After this TOF analysis of our detection system, we may display the energy and time-dependent measurements as 2D energy vs. time maps for the different water clusters, as shown in Figs. \ref{fig:2Dscan12} and \ref{fig:2Dscan34}. The 2D energy vs time maps also show the time-integrated signal on a side panel to the left, with the maximum marked by an orange line, and the energy-integrated signal on a panel below. The expected energy of a specific ion due to the supersonic expansion $E_0$ + $E_{\rm beam}$, obtained from eq. \ref{eq:ebeam}, is shown as orange dashed lines, the energy $E_0$ as a green dashed line. In the first attempt, all 2D energy vs. time maps are analysed using an energy scaling factor $c$ = 0.011 eV/amu and $E_0$ = -3.8 eV, as determined from the energy scans above. However, the 2D energy vs. time maps reveal distinct peaks in time and energy. Suppose we regard the most prominent peaks originating from supersonic expansion alone. The energy at the maximum intensity of those peaks for each cluster is used to extrapolate the offset $E_0$ and scaling factor $c$, leading to a more accurate estimation of $E_{\rm beam}$ for each mass. A scaling factor $c$ = 0.012 eV/amu and $E_0$ = -3.9 eV is obtained, which is in very good agreement with the $c$ value determined previously \cite{grosse-kreul_mass_2015}. In the following, we discuss these 2D energy vs. time maps of the different water clusters:

\begin{itemize}
\item \textbf{\waterone}: The 2D energy vs. time map for \waterone is shown in Fig. \ref{fig:2Dscan12}a. One can see that the strongest signal is at $E_{\rm beam}$ at $t$=10 $\mu$s. A small peak is observed at later times, with an energy corresponding to $E_{\rm beam}$ + 0.3 eV.

\item \textbf{\watertwo}: The 2D energy vs. time map for \watertwo is shown in Fig. \ref{fig:2Dscan12}b. A clear signal for \watertwo ions with an energy $E_{\rm beam}$ is observed. In addition, a smaller signal is also visible 10 $\mu$s later at an energy of $E_{\rm beam}$ + 0.3 eV. 

\begin{figure}
    \centering
\includegraphics[width=0.7\textwidth]{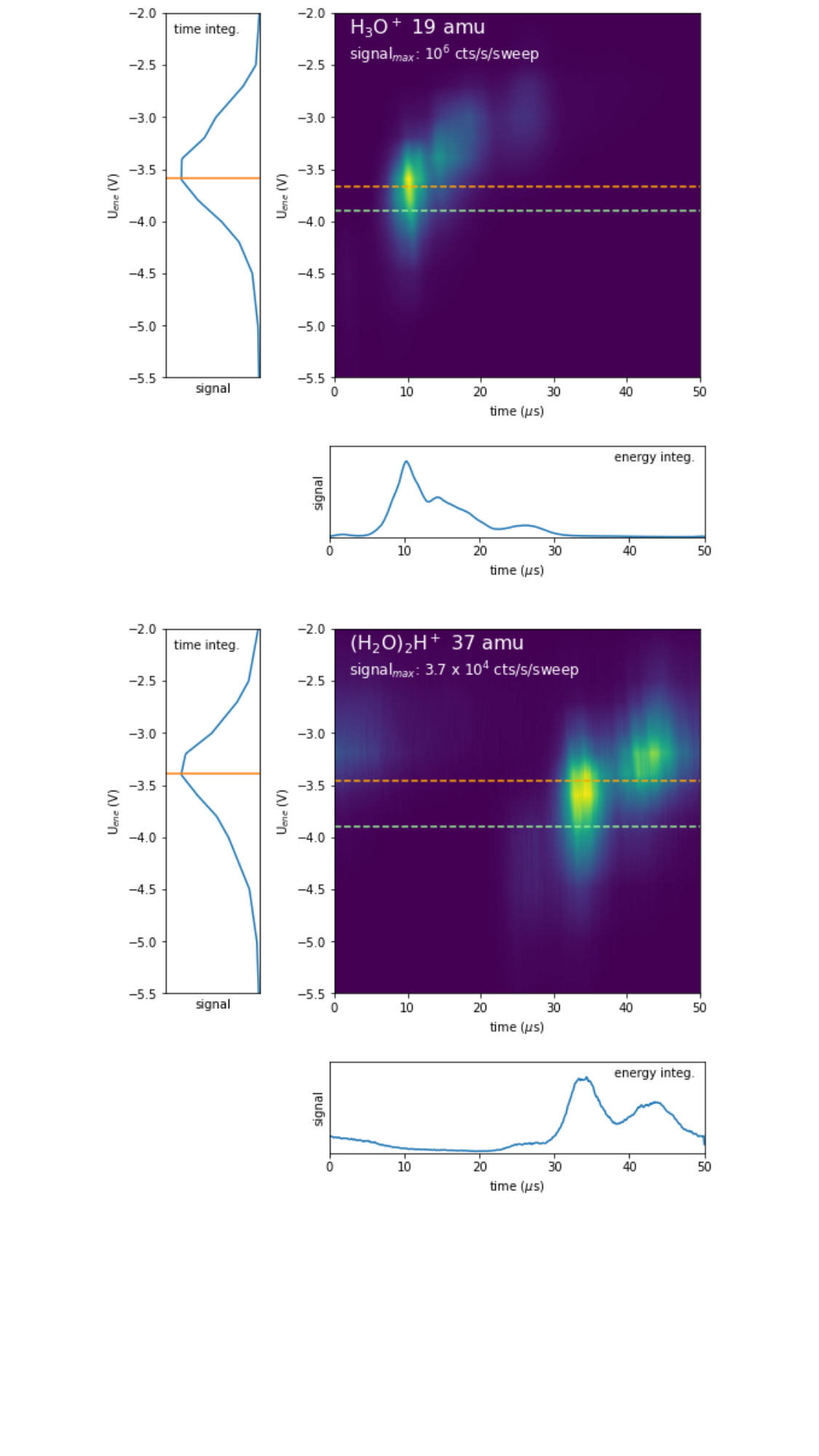}
    \caption{2D energy vs. time maps of the water cluster ions \waterone (top panel) and \watertwo (bottom panel). The small side panels show the time-integrated signals; the small panels below show the energy-integrated signals (same as Fig. \ref{fig:watercluster}). $E_0$ is shown as green dashed line, $E_0$ + $E_{\rm beam}$ as orange dashed line.}
    \label{fig:2Dscan12}
\end{figure}

\item \textbf{\waterthree}: The 2D energy vs. time map for \waterthree is shown in Fig. \ref{fig:2Dscan34}a. A clear signal for \waterthree ions with an energy of $E_{\rm beam}$ is observable. A faint signal is also seen 7 $\mu$s later at an energy of $E_{\rm beam}$ + 0.3 eV.

\item \textbf{\waterfour}: The 2D energy vs. time map for \waterfour is shown in Fig. \ref{fig:2Dscan34}b. A clear signal for \waterfour ions at an energy $E_{\rm beam}$ is observable. Again, a faint signal at 7 $\mu$s later is also seen at an energy of $E_{\rm beam}$ + 0.3 eV. Even a faint signal at 12 $\mu$s may be detected closer to $E_0$.

\begin{figure}
    \centering
\includegraphics[width=0.7\textwidth]{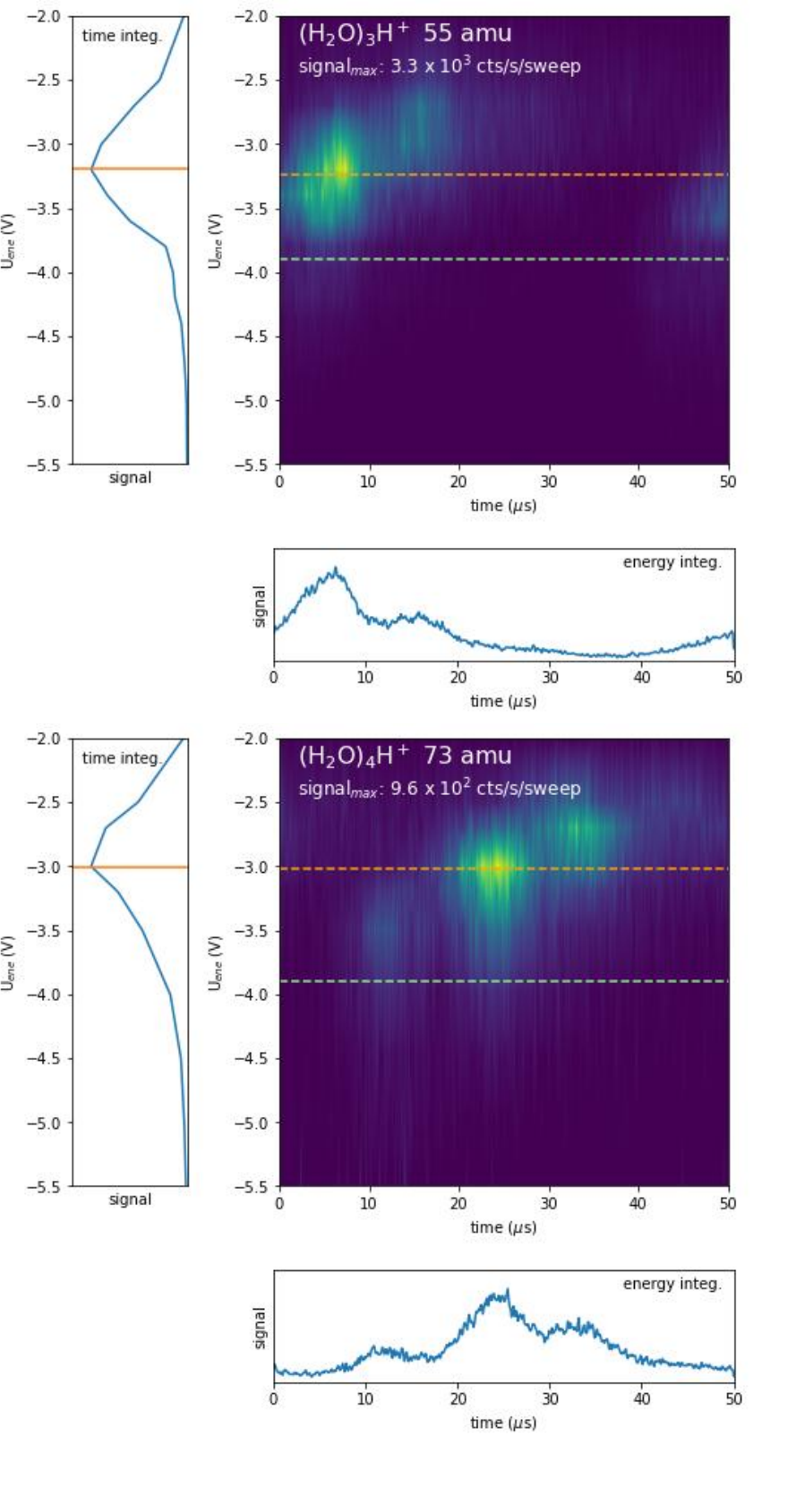}
    \caption{2D energy vs. time maps of the water cluster ions \waterthree (top panel) and \waterfour (bottom panel). The small side panels show the time-integrated signals; the small panels below show the energy-integrated signals (same as Fig. \ref{fig:watercluster}). $E_0$ is shown as green dashed line, $E_0$ + $E_{\rm beam}$ as orange dashed line.}
    \label{fig:2Dscan34}
\end{figure}
\end{itemize}

These 2D energy vs time maps of the different water clusters show that their energy increases with mass, as expected from the supersonic beam seeding. In addition, a much fainter signal is always seen 10 $\mu$s later at an energy of $E_{\rm beam}$ + 0.3 eV. One exception is a faint signal for \waterfour with an energy close to $E_0$. 

In all 2D energy vs. time maps, a reconstructed energy scan, likewise shown in figure \ref{fig:energywatercluster}, is also shown on the left panels with a maximum indicated by an orange solid line. One can see that this averaging over the time axis yields a different maximum of the energy scan than the $E_{beam}$ values determined from the prominent peak of the temporally resolved measurement (dashed orange line). This holds especially for \waterone and \watertwo, where the averaging of the signals from different peaks of the accelerated ions yields larger values than $E_{beam}$. This explains the different scaling factors of $c$ = 0.011 eV/amu for the simple energy scan versus $c$ = 0.012 eV/amu for the 2D energy vs. time maps. Both scaling factors are not contradictory but reflect that more detailed data are obtained by analysing the time-dependence of the signals.

\newpage
\clearpage

\subsubsection{Nitrogen and oxygen ions}~\\

\paragraph{Energy vs. time maps of the impacting nitrogen and oxygen ions}~\\

The analysis of the water cluster spectra revealed a consistent picture, where all water cluster ions exhibit an energy $E_{\rm beam}$ originating from the supersonic expansion and a characteristic higher energy by 0.3 eV at later times. In the following, we regard nitrogen- and oxygen-containing ions. As a reference, we use the energy scaling determined from the water cluster ions with the corresponding mass-dependent values for $E_{\rm beam}$ and the reference energy $E_0$. The 2D energy vs. time maps are discussed in the following. 

\begin{itemize}
\item\textbf{N$_2^+$:} Fig. \ref{fig:2DscanN2NO}a shows the 2D energy vs. time map of N$_2^+$ ions. A clear signal for N$_2^+$ ions with an energy at $E_{\rm beam}$ is observable. Also, a faint signal at 10 $\mu$s later at an energy of 0.3 eV above $E_{\rm beam}$ is visible. A faint signal at energies below $E_0$ can be identified at $t$ = 18 $\mu$s.

\item\textbf{NO$^+$:} Fig. \ref{fig:2DscanN2NO}b shows the 2D energy vs. time map of NO$^+$ ions. A signal for NO$^+$ ions at an energy almost at $E_{\rm beam}$ is observable. Also, a faint signal is visible 10 $\mu$s later at an energy of 0.3 eV above $E_{\rm beam}$. Also, a clear signal for NO$^+$ at energies above $E_{\rm beam}$ is clearly visible, with energy slowly increasing over time.

\begin{figure}
    \centering
\includegraphics[width=0.7\textwidth]{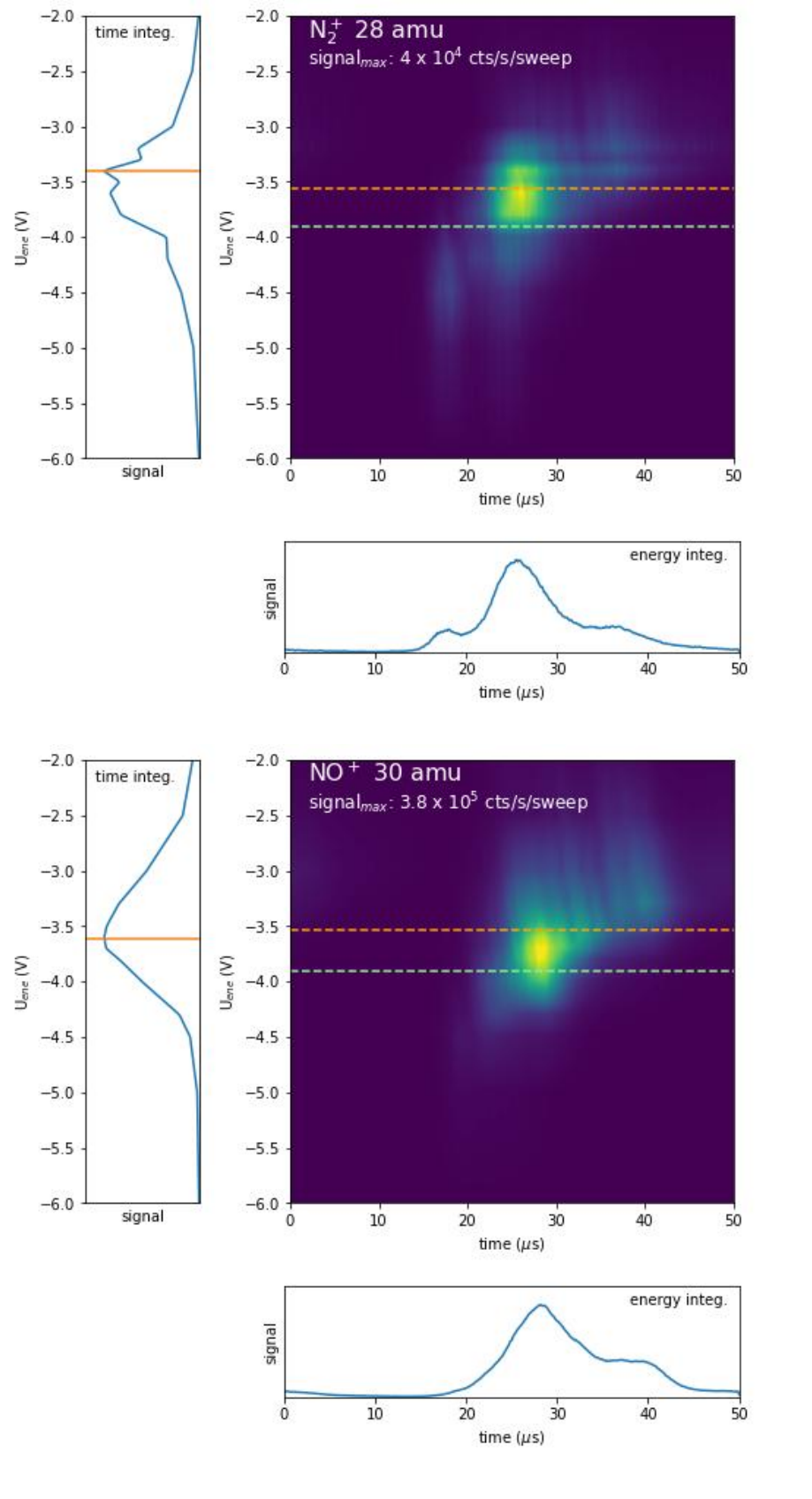}
    \caption{2D energy vs. time maps of N$_2^+$ ions (top panel) and NO$^+$ (bottom panel). The small side panels show the time-integrated signals; the small panels below show the energy-integrated signals. $E_0$ is shown as green dashed line, $E_0$ + $E_{\rm beam}$ as orange dashed line.}
    \label{fig:2DscanN2NO}
\end{figure}

\item\textbf{O$_2^+$:} Fig. \ref{fig:2DscanO2O3}a shows the 2D energy vs. time map of O$_2^+$ ions. A strong signal for O$_2^+$ ions at an energy $E_{\rm beam}$ is observed. Also, a clear signal for O$_2^+$ at energies above $E_{\rm beam}$ is clearly visible, with energy slowly increasing over time.  

\item\textbf{O$_3^+$ (or (H$_2$O)NO$^+$):} Fig. \ref{fig:2DscanO2O3}b shows the 2D energy vs. time map of O$_3^+$ (or (H$_2$O)NO$^+$) ions. Due to the identical mass of 48 amu, it is impossible to decide whether the signal originates from O$_3^+$ or (H$_2$O)NO$^+$ ions. In the following, we refer to O$_3^+$ only for simplicity. A clear signal for O$_3^+$ ions at an energy $E_{\rm beam}$ is observed. Also, a faint signal at 0.3 eV above $E_{\rm beam}$ might be detected at later times, and a faint signal increasing from $E_0$ to $E_{\rm beam}$ at earlier times.  

\begin{figure}
    \centering
\includegraphics[width=0.7\textwidth]{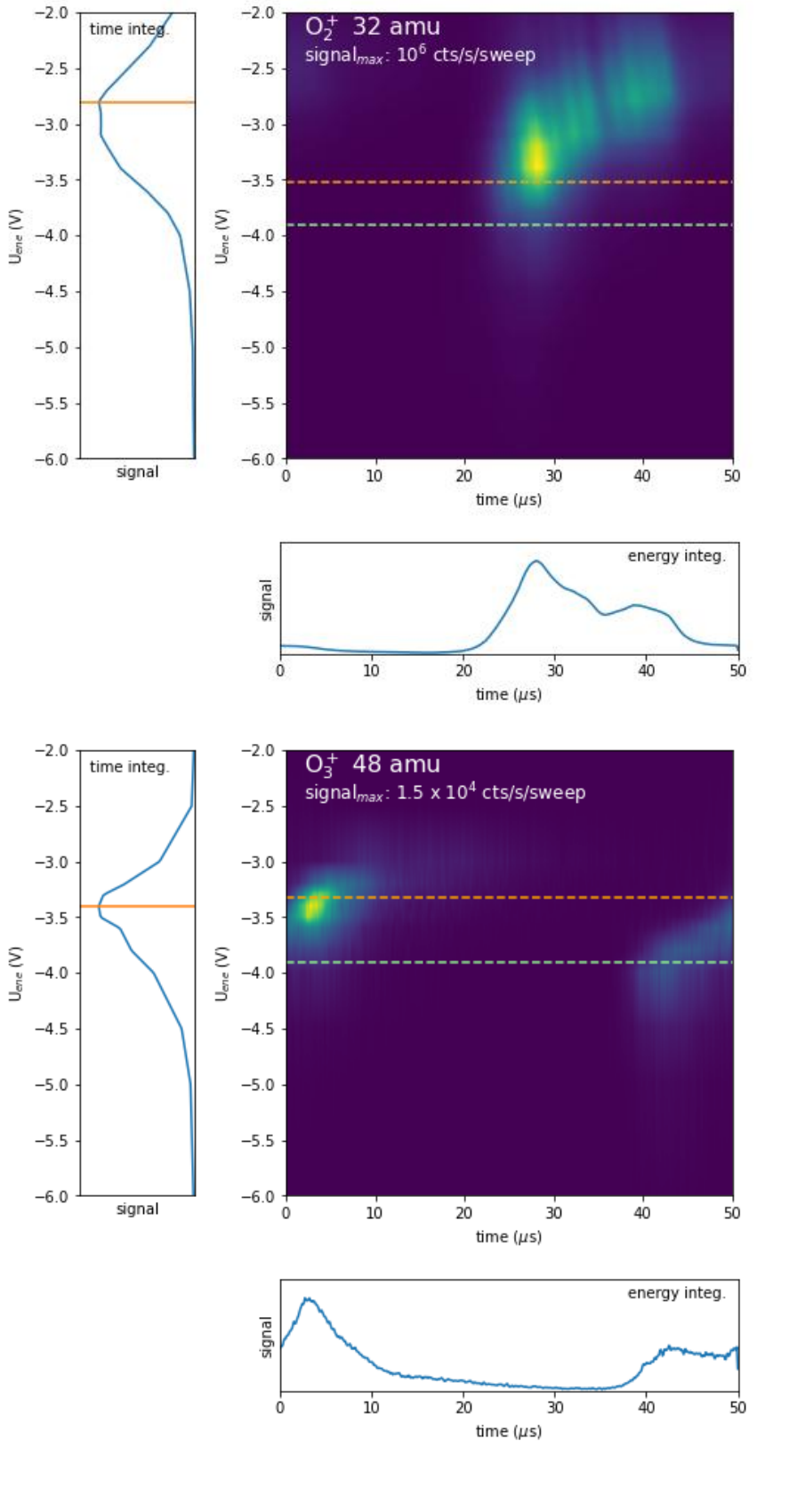}
    \caption{2D energy vs. time maps of O$_2^+$ ions (top panel) and O$_3^+$ (bottom panel). The small side panels show the time-integrated signals; the small panels below show the energy-integrated signals. $E_0$ is shown as a green dashed line, $E_0$ + $E_{\rm beam}$ as an orange dashed line.}   
    \label{fig:2DscanO2O3}
\end{figure}
\end{itemize}

The analysis of nitrogen and oxygen-containing ions shows that for N$_2^+$, NO, O$_2^+$ and O$_3^+$ ions, peaks at $E_{\rm beam}$ are seen, similar to the energy scaling of the water cluster ions. However, a closer inspection also indicates that the energy of O$_2^+$ is slightly above that of NO$^+$ and N$_2^+$ by 0.5 eV. Such a difference cannot originate from the supersonic expansion's mass scaling because these three ions' atomic masses are very close. This implies that the chemical identity of the ions has a slight influence on their absolute energy.

This interpretation of the 2D energy vs. time maps for water clusters and nitrogen and oxygen ions is limited in accuracy, given the sampling process's complexity and the spectrometer's finite resolution. Nevertheless, we discuss a consistent picture to explain the energy and time scale of all observed peaks in the following.   

\subsection{Comparison metallic vs. dielectric orifice}

\begin{figure}
    \centering
\includegraphics[width=0.7\textwidth]{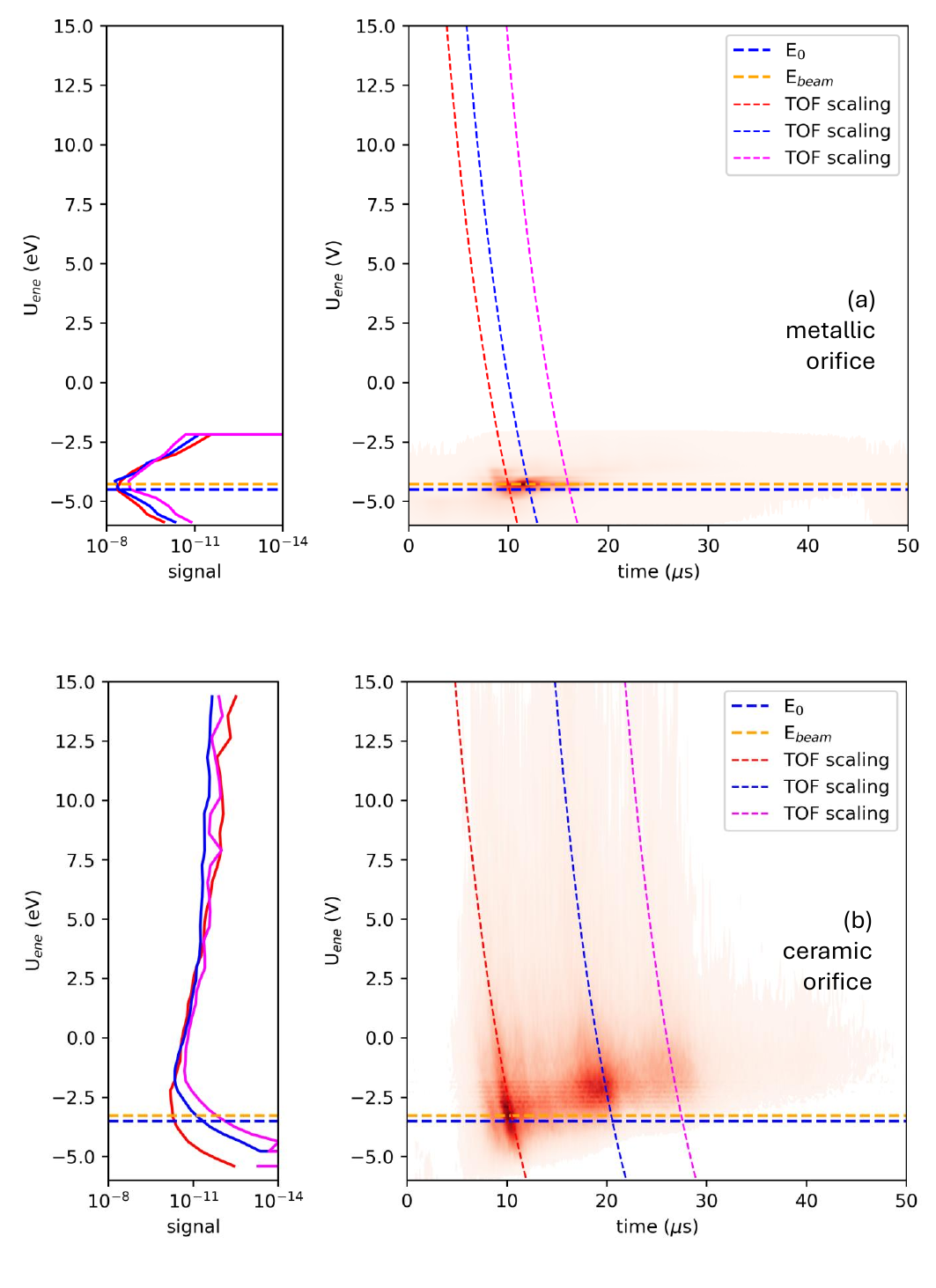}
    \caption{2D energy vs. time maps of \waterone ions, sampled via a metallic orifice (a) and a dielectric orifice (b). The data for the metallic orifice are collected up to $U_{\rm ene}$ = -2 eV and for the ceramic orifice up to $U_{\rm ene}$ = 15 eV. The dashed colour lines illustrate the TOF scaling following $U_{ene} \propto 1/t^2$. The left panel shows the energy distributions for the ions at different times.}
    \label{fig:ceramic}
\end{figure}

In a separate experimental campaign, the metallic orifice was replaced by a ceramic orifice and the 2D energy vs time maps were acquired for water cluster ions. An example is shown in Fig. \ref{fig:ceramic} for \waterone comparing the metallic (a) with the ceramic (b) orifice. The left panels show the ion energy distribution function in counts per sweep at individual times. For this, the TOF of the ions is accounted for. First we considered the scaling of the TOFs following eq. \ref{eq:tofscaling}, using a constant mass $m$ but varying $E$. Since the TOF differences for varying energy are accumulated in the energy filter, a smaller travel distance $d$ has to be used. In the end, a fit of TOF directly from the measured 2D maps is performed as $U_{\rm ene} = k/(t-t_0)^2$ with a scaling constant $k$ and an offset $t_0$ to create temporal isolines extending from lower to higher energies. Then, the IEDFs are extracted along the isolines connecting the signal of ions that enter the orifice simultaneously but are detected at different times by the MBMS. The ions sampled from the metallic orifice exhibit a much higher intensity but with energies much smaller. An energy scan up to $U_{\rm ene}$=-2 eV was sufficient. 
The reason for the drastically increased energy range observed when sampling with a dielectric orifice remains unclear. The high energy ions seen in the simulations of Babaeva and Kushner \cite{babaeva_ion_2011,babaeva_ion_2011a,babaeva_control_2013} only appear during the lifetime of the initial streamer, while the measured high energy ions remain long after the bullets arrival. In an additional experiment (see the supplementary material), a 25 $\mu$m dielectric orifice was used to measure the IEDF of H$_3$O$^+$. The resulting IEDF does not display significant differences to the one obtained using the 50 $\mu$m orifice in terms of energy range. Consequently, these ions are unlikely to originate from the streamer penetrating into the orifice.

\newpage
\clearpage

\section{Discussion}

%\subsection{Processes that affect the energy of the sampled ions}

Analysing the ion energy spectrum from the dielectric barrier discharge plasma jet is an intricate challenge since many processes may affect the energy of the sampled ions, either directly from the plasma or by processes inside the mass spectrometer.

\begin{itemize}
\item\textit{Supersonic expansion:} The supersonic expansion accelerates all species to the same velocity, as described in the experimental section above. This occurs up to a position $z_{\rm quitting}$ = 60 $\mu$m before the beam enters the molecular regime. Any collisions inside this expansion region might alter the identity of the accelerated ions, but the final velocity remains unchanged. Therefore, collisions within the supersonic expansion do not affect the energy scaling of the particles. All ions exhibit the same velocity but different energy at the quitting surface according to $E_{\rm beam} = c m$. The values for $E_{beam}$ range from 0.23 eV to 0.88 eV for the smallest to the largest water cluster ion, assuming a scaling factor $c=0.012$. Nevertheless, any chemical reactions inside the expanding beam may alter the composition of the sampled species. For example, larger water clusters grow during the expansion and cooling of the beam. The energies of the sampled ions can be lower than $E_{\rm beam}$ if the momentum transfer of the sampled ions to the carrier gas is incomplete. This is referred to as the so-called slipping in a seeded beam \cite{Morse_1996}.

\item\textit{Plasma afterglow:} After the streamer head impacts the surface, we observed a long-lasting emission at the orifice position over 1.5 $\mu$s corresponding to the length of the positive voltage pulse. After the streamer propagation stops, ambipolar fields are expected to create a quasineutral region rather quickly before electron-ion recombination sets in with a decay constant of typically 100 ns\cite{orriere_ionization_2018,zauner-wieczorek_ion_2022}. However, the analysis of streamers in front of surfaces showed that a sizeable positive space charge region prevails because electrons are lost to the surface during streamer propagation\cite{kourtzanidis_electrohydrodynamic_2021,kourtzanidis_selfconsistent_2021}. This positive space charge region in front of the surface can only decay on a time scale of the ion diffusion to the surface. Assuming a typical diffusion constant of $D$ = 6 $\times$ 10$^{-5}$ m$^2$/s for N$_2$ in He \cite{wasik_measurements_1969} and a travel distance of $d$ = 30 $\mu$m to the orifice, we obtain a travel time $t = d^2/D$ = 15 $\mu$s. This is in the range of the observed trends. Therefore, we propose the following: the impact of the positive streamer creates an intense cloud of electrons and ions in front of the orifice, but the electrons are swiftly lost to the surface, and a residual positive charge remains at the surface, as well as a positive space-charge region in front of the surface. Now, the positive ions diffuse to the surface over 10 $\mu$s. The surface charge increases further during this time, causing a slow shift in $E_0$ to higher values. This process is the same for all ions. 

%The electron density is expected to decrease over time after the streamer head impact over this 1.5 $\mu$s. Even at later times, ion-molecule reactions may lead to larger molecular ions, such as the creation of water cluster ions, but also the creation of ozone ions in reactions of O$_2^+$ with O (or O$^+$ with O$_2$) to form O$_3^+$ as well as nitric oxide from N$^+$ reacting with O$_3$ creating NO$^+$ + O$_2$. All these secondary ions are collected by the gas stream entering the orifice. 

The chemistry in the decaying space-charge region in front of the surface also leads to the slow growth of water clusters. The same sequence can also be seen in the oxygen- and nitrogen-containing ions. Oxygen, nitrogen, and NO$^+$ ions are observed first before the impact of O$_3^+$ becomes visible. Since O$_3^+$ is only formed in the plasma afterglow, these species are not expected during the impact of the streamer head. In accordance with that, the time of flight of ozone ions predicted in figure \ref{fig:tofsim} is noticeably smaller than the appearance time at the detector.

\item\textit{Surface charges:} During the plasma impact, $E_0$ may also vary because the surface charge may change. For example, if the surface charges positively, the potential energy of an ion in front of the surface also increases, eventually shifting the energies to higher values. This charge-induced change of the $E_0$ may also explain peaks that appear below $E_0$ in some of the 2d energy vs. time maps. It is conceivable that the surface undergoes a period cycle of being charged slightly negatively at first by electron impact before being charged positively by the incident ions. In all spectra, the IEDFs start at small values for $E_0$ earlier, with $E_0$ increasing similarly over time in a characteristic manner for all ions. This is observed in the data.
\end{itemize}

All of these effects cause a shift in the observed energy of the ion. However, one may distinguish effects that may cause a distribution of energies ranging from $E_{\rm beam}$  to lower or higher energies and effects that may cause peaks in the energy vs. time map that are distinct from $E_{\rm beam}$. For example, Penning ionisation in the expanding beam or injection of plasma ions via collisional sheaths will lead to distributions in energy ending or starting at $E_{\rm beam}$. In contrast, ionic work function shifts or surface charging will lead to distinct peaks that are shifted with respect to $E_{\rm beam}$ assuming an uncharged surface, for example. As the observed energies do not differ significantly from the energies dictated by supersonic expansion, the penetration of the plasma into the orifice, while possible, is unlikely to affect the measured IEDFs in this case (see also He$^+$ energy scan in supplementary material).

Applying this to the measured 2D energy vs time maps of water clusters, the high-intensity peak can be attributed to the ions initially formed during the impact of the plasma bullet, as they appear at the energy corresponding to the supersonic expanding beam. The ions from the decaying space charge region in front of the orifice appear at higher energies at later times due to the continued charging of the surface.

Next, we regard nitrogen- and oxygen-containing ions. At first, it can be noted that the values for $E_{beam}$ denoted from water cluster ions do not perfectly match the prominent peak of the NO$_x$ ions. This could be caused by different synthesis pathways of NO$_x$ ions being created more directly in the plasma, whereas most of the water ions are created in the plasma afterglow. However, the chemical identity of the ions also affects the absolute energy scale due to their ionic work function: the actual kinetic energy of sampled species depends on the energy calibration of the MBMS, which is in reference to the Fermi energy difference of the SEM versus the orifice material (this is equivalent to any electron spectroscopies such as photoelectron spectroscopy (XPS), where the energy scale is referenced to the electronic work function of the entrance orifice of the electron energy analyser). Since the potential energy of an ion differs in front of a surface, the energy scale also varies slightly, by a few 0.1 eV at most. This is beyond the scope of this paper. Apart from these minute differences, the general trends measured in NO$_x$ species are congruent to those seen with the water cluster. An exception is the temporal behaviour of O$_3^+$, as it is more readily created during the afterglow phase of the plasma.

To conclude, one can state that the IEDFs of all ions are dominated by seeding in the supersonic expanding helium beam into the MBMS. No high energy ions, as predicted by Babeava and Kushner have been found. This cannot exclude their existence, since they should appear only in a ns time window, where most of the ions are sampled over many microseconds. The energy scale of the incident ion is significantly influenced by the charging of the surface by electrons and positive ions. The ion flux to the surface lasts over many microseconds, much longer than the impact of the streamer, but also much longer than the width of the voltage pulse. This is explained by a positive space charge region in front of the surface, which slowly decays over time. The effect of surface charging is more significant in the case of a ceramic material, as expected. A better description of such a decaying space charge in front of the surface requires plasma models that cover the actual plasma phase and the interplay between residual charges and surface states on a much longer time scale.

\newpage
\clearpage

\section{Conclusion}

Time-resolved molecular beam mass spectrometry with an energy filter has been used to analyse the energy spectra and the temporal evolution of ions generated by the impact of a dielectric barrier discharge plasma jet at a surface. It is shown that all ions are sampled at energies below 1 eV since the plasma sheaths are collisional, so any high-energy ions lose their energy in collisions before impact. After the plasma impact, the surface is positively charged, so the MBMS collects ions at a slightly higher energy of 0.3 eV. This occurs in an afterglow created by a positive space-charge region in front of the surface. The temporal sequence of the ions is consistent with the known chemistry of air plasmas. These measurements successfully monitor the ion chemistry and the charging process of a streamer interacting with a surface. Sampling ions at energies of a few eV only by MBMS is challenging because beam seeding may destroy the information on the initial energy of the ions, since all ions appear at $E_{beam}$ in the mass spectrometer. An alternative to monitor the IEDFs would be laser spectroscopy methods such as laser-induced fluorescence (LIF). This is also challenging due to the small kinetic energies of the species and the necessity to observe them very close to a surface, where any reflections spoil the signals. Ultimately, a more thorough analysis of the MBMS data would require a dedicated model that includes the plasma phase, the plasma afterglow, and the complete sampling process after the interplay between streamer plasmas and surfaces.

\section*{Ethical Approval} 
not applicable

\section*{Competing interests}

No conflict of interest on behalf of all authors.
 
\section*{Authors' contributions}

\section*{Acknowledgment} 

The work is supported by the SFB1316, funded by the German Science Foundation.

\section*{Availability of data and materials} 
The authors can provide all data upon reasonable request.

\printbibliography

@article{mohsenimehr_plasma_2025,
	title = {Plasma and Flow Simulation of the Ion Wind in a Surface Barrier Discharge Used for Gas Conversion Benchmarked by Schlieren Imaging},
	volume = {45},
	issn = {0272-4324, 1572-8986},
	url = {https://link.springer.com/10.1007/s11090-024-10533-0},
	doi = {10.1007/s11090-024-10533-0},
	pages = {85--112},
	number = {1},
	journaltitle = {Plasma Chemistry and Plasma Processing},
	shortjournal = {Plasma Chem Plasma Process},
	author = {Mohsenimehr, S. and Wilczek, S. and Mussenbrock, T. and Keudell, A. Von},
	urldate = {2025-03-24},
	date = {2025-01},
	langid = {english},
}

@article{aleksandrov_simulation_1996,
  title = {Simulation of Long-Streamer Propagation in Air at Atmospheric Pressure},
  author = {Aleksandrov, N L and Bazelyan, E M},
  year = {1996},
  month = mar,
  journal = {Journal of Physics D: Applied Physics},
  volume = {29},
  number = {3},
  pages = {740--752},
  issn = {0022-3727, 1361-6463},
  doi = {10.1088/0022-3727/29/3/035},
  urldate = {2024-11-17},
  langid = {english}
}

@article{babaeva_control_2013,
  title = {Control of Ion Activation Energy Delivered to Tissue and Sensitive Materials in Atmospheric Pressure Plasmas Using Thin Porous Dielectric Sheets},
  author = {Babaeva, Natalia Yu and Kushner, Mark J},
  year = {2013},
  month = mar,
  journal = {Journal of Physics D: Applied Physics},
  volume = {46},
  number = {12},
  pages = {125201},
  issn = {0022-3727, 1361-6463},
  doi = {10.1088/0022-3727/46/12/125201},
  urldate = {2024-11-16},
  copyright = {http://iopscience.iop.org/info/page/text-and-data-mining},
  langid = {english}
}

@article{babaeva_ion_2011,
  title = {Ion Energy and Angular Distributions onto Polymer Surfaces Delivered by Dielectric Barrier Discharge Filaments in Air: {{II}}. {{Particles}}},
  shorttitle = {Ion Energy and Angular Distributions onto Polymer Surfaces Delivered by Dielectric Barrier Discharge Filaments in Air},
  author = {Babaeva, Natalia Yu and Kushner, Mark J},
  year = {2011},
  month = jun,
  journal = {Plasma Sources Science and Technology},
  volume = {20},
  number = {3},
  pages = {035018},
  issn = {0963-0252, 1361-6595},
  doi = {10.1088/0963-0252/20/3/035018},
  urldate = {2024-11-16},
  langid = {english}
}

@article{babaeva_ion_2011a,
  title = {Ion Energy and Angular Distributions onto Polymer Surfaces Delivered by Dielectric Barrier Discharge Filaments in Air: {{I}}. {{Flat}} Surfaces},
  shorttitle = {Ion Energy and Angular Distributions onto Polymer Surfaces Delivered by Dielectric Barrier Discharge Filaments in Air},
  author = {Babaeva, Natalia Yu and Kushner, Mark J},
  year = {2011},
  month = jun,
  journal = {Plasma Sources Science and Technology},
  volume = {20},
  number = {3},
  pages = {035017},
  issn = {0963-0252, 1361-6595},
  doi = {10.1088/0963-0252/20/3/035017},
  urldate = {2024-11-17},
  langid = {english}
}

@article{babaeva_plasma_2019,
  title = {Plasma Bullet Propagation and Reflection from Metallic and Dielectric Targets},
  author = {Babaeva, Natalia Yu and Naidis, George V and Panov, Vladislav A and Wang, Ruixue and Zhang, Shuai and Zhang, Cheng and Shao, Tao},
  year = {2019},
  month = sep,
  journal = {Plasma Sources Science and Technology},
  volume = {28},
  number = {9},
  pages = {095006},
  issn = {1361-6595},
  doi = {10.1088/1361-6595/ab36d3},
  urldate = {2024-11-18},
  langid = {english}
}

@article{chauvet_characterization_2014,
  title = {Characterization of an Asymmetric {{DBD}} Plasma Jet Source at Atmospheric Pressure},
  author = {Chauvet, Laura and Th{\'e}r{\`e}se, Laurent and Caillier, Bruno and Guillot, Philippe},
  year = {2014},
  month = oct,
  journal = {J. Anal. At. Spectrom.},
  volume = {29},
  number = {11},
  pages = {2050--2057},
  issn = {0267-9477, 1364-5544},
  doi = {10.1039/C4JA00255E},
  urldate = {2024-11-17},
  langid = {english}
}

@article{grosse-kreul_mass_2015,
  title = {Mass Spectrometry of Atmospheric Pressure Plasmas},
  author = {{Gro{\ss}e-Kreul}, S and H{\"u}bner, S and Schneider, S. and Ellerweg, D and {von Keudell}, A and Matej{\v c}{\'i}k, S and Benedikt, J},
  year = {2015},
  month = jul,
  journal = {Plasma Sources Science and Technology},
  volume = {24},
  number = {4},
  pages = {044008},
  publisher = {IOP Publishing},
  doi = {10.1088/0963-0252/24/4/044008}
}

@phdthesis{GrosseKreul_phd,
  type = {Doctoralthesis},
  title = {Mass Spectrometry of Ions from Atmospheric Pressure Plasmas},
  author = {{Gro{\ss}e-Kreul}, Simon},
  year = {2016},
  school = {Ruhr-Universit{\"a}t Bochum, Universit{\"a}tsbibliothek}
}

@article{guaitella_impingement_2015,
  title = {The Impingement of a {{kHz}} Helium Atmospheric Pressure Plasma Jet on a Dielectric Surface},
  author = {Guaitella, O and Sobota, A},
  year = {2015},
  month = jun,
  journal = {Journal of Physics D: Applied Physics},
  volume = {48},
  number = {25},
  pages = {255202},
  issn = {0022-3727, 1361-6463},
  doi = {10.1088/0022-3727/48/25/255202},
  urldate = {2024-11-18},
  langid = {english}
}

@article{huang_thermodynamic_2018,
  title = {Thermodynamic {{Model}} for {{Calculation}} the {{Position}} of {{Quitting Surface}} of {{Supersonic Beam Based}} on {{Speed Measurements}}},
  author = {Huang, Chuanfu},
  year = {2018},
  month = nov,
  journal = {The Journal of Physical Chemistry A},
  volume = {122},
  number = {46},
  pages = {8998--9000},
  issn = {1089-5639, 1520-5215},
  doi = {10.1021/acs.jpca.8b09149},
  urldate = {2024-11-01},
  langid = {english}
}

@article{ito_mass_2015,
  title = {Mass {{Spectrometry Analyses}} of {{Ions Generated}} by {{Atmospheric-Pressure Plasma Jets}} in {{Ambient Air}}},
  author = {Ito, Tomoko and Gotoh, Kensaku and Sekimoto, Kanako and Hamaguchi, Satoshi},
  year = {2015},
  journal = {Plasma Medicine},
  volume = {5},
  number = {2-4},
  pages = {283--298},
  issn = {1947-5764},
  doi = {10.1615/PlasmaMed.2016016443},
  urldate = {2024-11-18},
  langid = {english}
}

@article{jarrige_formation_2010,
  title = {Formation and Dynamics of Plasma Bullets in a Non-Thermal Plasma Jet: Influence of the High-Voltage Parameters on the Plume Characteristics},
  shorttitle = {Formation and Dynamics of Plasma Bullets in a Non-Thermal Plasma Jet},
  author = {Jarrige, Julien and Laroussi, Mounir and Karakas, Erdinc},
  year = {2010},
  month = dec,
  journal = {Plasma Sources Science and Technology},
  volume = {19},
  number = {6},
  pages = {065005},
  issn = {0963-0252, 1361-6595},
  doi = {10.1088/0963-0252/19/6/065005},
  urldate = {2024-11-18},
  langid = {english}
}

@article{jiang_absolute_2021,
  title = {Absolute Ion Density Measurements in the Afterglow of a Radiofrequency Atmospheric Pressure Plasma Jet},
  author = {Jiang, Jingkai and Bruggeman, Peter J},
  year = {2021},
  month = apr,
  journal = {Journal of Physics D: Applied Physics},
  volume = {54},
  number = {15},
  pages = {15LT01},
  issn = {0022-3727, 1361-6463},
  doi = {10.1088/1361-6463/abdc91},
  urldate = {2024-11-17},
  langid = {english}
}

@article{klarenaar_how_2018,
  title = {How Dielectric, Metallic and Liquid Targets Influence the Evolution of Electron Properties in a Pulsed {{He}} Jet Measured by {{Thomson}} and {{Raman}} Scattering},
  author = {Klarenaar, B L M and Guaitella, O and Engeln, R and Sobota, A},
  year = {2018},
  month = aug,
  journal = {Plasma Sources Science and Technology},
  volume = {27},
  number = {8},
  pages = {085004},
  issn = {1361-6595},
  doi = {10.1088/1361-6595/aad4d7},
  urldate = {2024-11-18},
  langid = {english}
}

@article{kondeti_longlived_2018,
  title = {Long-Lived and Short-Lived Reactive Species Produced by a Cold Atmospheric Pressure Plasma Jet for the Inactivation of {{Pseudomonas}} Aeruginosa and {{Staphylococcus}} Aureus},
  author = {Kondeti, V.S. Santosh K. and Phan, Chi Q. and Wende, Kristian and Jablonowski, Helena and Gangal, Urvashi and Granick, Jennifer L. and Hunter, Ryan C. and Bruggeman, Peter J.},
  year = {2018},
  month = aug,
  journal = {Free Radical Biology and Medicine},
  volume = {124},
  pages = {275--287},
  issn = {08915849},
  doi = {10.1016/j.freeradbiomed.2018.05.083},
  urldate = {2024-11-17},
  langid = {english}
}

@article{kourtzanidis_electrohydrodynamic_2021,
  title = {The Electrohydrodynamic Force Distribution in Surface {{AC}} Dielectric Barrier Discharge Actuators: Do Streamers Dictate the Ionic Wind Profiles?},
  shorttitle = {The Electrohydrodynamic Force Distribution in Surface {{AC}} Dielectric Barrier Discharge Actuators},
  author = {Kourtzanidis, K and Dufour, G and Rogier, F},
  year = {2021},
  month = jul,
  journal = {Journal of Physics D: Applied Physics},
  volume = {54},
  number = {26},
  pages = {26LT01},
  issn = {0022-3727, 1361-6463},
  doi = {10.1088/1361-6463/abf53e},
  urldate = {2023-07-13},
  langid = {english}
}

@article{kourtzanidis_selfconsistent_2021,
  title = {Self-Consistent Modeling of a Surface {{AC}} Dielectric Barrier Discharge Actuator: {{In-depth}} Analysis of Positive and Negative Phases},
  shorttitle = {Self-Consistent Modeling of a Surface {{AC}} Dielectric Barrier Discharge Actuator},
  author = {Kourtzanidis, K and Dufour, G and Rogier, F},
  year = {2021},
  month = jan,
  journal = {Journal of Physics D: Applied Physics},
  volume = {54},
  number = {4},
  pages = {045203},
  issn = {0022-3727, 1361-6463},
  doi = {10.1088/1361-6463/abbcfd},
  urldate = {2023-07-13},
  langid = {english}
}

@article{Lu_2014,
  title = {Guided Ionization Waves: {{Theory}} and Experiments},
  author = {Lu, X. and Naidis, G.V. and Laroussi, M. and Ostrikov, K.},
  year = {2014},
  journal = {Physics Reports},
  volume = {540},
  number = {3},
  pages = {123--166},
  issn = {0370-1573},
  doi = {10.1016/j.physrep.2014.02.006}
}

@article{lu_atmosphericpressure_2012,
  title = {On Atmospheric-Pressure Non-Equilibrium Plasma Jets and Plasma Bullets},
  author = {Lu, X and Laroussi, M and Puech, V},
  year = {2012},
  month = jun,
  journal = {Plasma Sources Science and Technology},
  volume = {21},
  number = {3},
  pages = {034005},
  issn = {0963-0252, 1361-6595},
  doi = {10.1088/0963-0252/21/3/034005},
  urldate = {2024-11-17},
  langid = {english}
}

@article{mericam-bourdet_experimental_2009,
  title = {Experimental Investigations of Plasma Bullets},
  author = {{Mericam-Bourdet}, N and Laroussi, M and Begum, A and Karakas, E},
  year = {2009},
  month = mar,
  journal = {Journal of Physics D: Applied Physics},
  volume = {42},
  number = {5},
  pages = {055207},
  issn = {0022-3727, 1361-6463},
  doi = {10.1088/0022-3727/42/5/055207},
  urldate = {2024-11-18},
  langid = {english}
}

@article{Morse_1996,
  title = {2. {{Supersonic}} Beam Sources},
  author = {Morse, Michael D.},
  year = {1996},
  month = jan,
  journal = {Experimental Methods in the Physical Sciences},
  volume = {29},
  pages = {21--47},
  doi = {10.1016/S0076-695X(08)60784-X}
}

@article{naidis_production_2014,
  title = {Production of Active Species in Cold Helium--Air Plasma Jets},
  author = {Naidis, G V},
  year = {2014},
  month = sep,
  journal = {Plasma Sources Science and Technology},
  volume = {23},
  number = {6},
  pages = {065014},
  issn = {0963-0252, 1361-6595},
  doi = {10.1088/0963-0252/23/6/065014},
  urldate = {2024-11-18},
  copyright = {http://iopscience.iop.org/info/page/text-and-data-mining},
  langid = {english}
}

@article{nijdam_physics_2020,
  title = {The Physics of Streamer Discharge Phenomena},
  author = {Nijdam, Sander and Teunissen, Jannis and Ebert, Ute},
  year = {2020},
  month = nov,
  journal = {Plasma Sources Science and Technology},
  volume = {29},
  number = {10},
  pages = {103001},
  issn = {1361-6595},
  doi = {10.1088/1361-6595/abaa05},
  urldate = {2024-11-18},
  langid = {english}
}

@article{oh_investigating_2015,
  title = {Investigating the Effect of Additional Gases in an Atmospheric-Pressure Helium Plasma Jet Using Ambient Mass Spectrometry},
  author = {Oh, Jun-Seok and Furuta, Hiroshi and Hatta, Akimitsu and Bradley, James W.},
  year = {2015},
  month = jan,
  journal = {Japanese Journal of Applied Physics},
  volume = {54},
  number = {1S},
  pages = {01AA03},
  issn = {0021-4922, 1347-4065},
  doi = {10.7567/JJAP.54.01AA03},
  urldate = {2024-11-17},
  copyright = {http://iopscience.iop.org/info/page/text-and-data-mining},
  langid = {english}
}

@article{orriere_ionization_2018,
  title = {Ionization and Recombination in Nanosecond Repetitively Pulsed Microplasmas in Air at Atmospheric Pressure},
  author = {Orri{\`e}re, Thomas and Moreau, Eric and Pai, David Z},
  year = {2018},
  month = dec,
  journal = {Journal of Physics D: Applied Physics},
  volume = {51},
  number = {49},
  pages = {494002},
  issn = {0022-3727, 1361-6463},
  doi = {10.1088/1361-6463/aae134},
  urldate = {2024-07-07},
  langid = {english}
}

@article{robert_experimental_2009,
  title = {Experimental {{Study}} of a {{Compact Nanosecond Plasma Gun}}},
  author = {Robert, Eric and Barbosa, Emerson and Dozias, S{\'e}bastien and Vandamme, Marc and Cachoncinlle, Christophe and Viladrosa, Raymond and Pouvesle, Jean Michel},
  year = {2009},
  month = dec,
  journal = {Plasma Processes and Polymers},
  volume = {6},
  number = {12},
  pages = {795--802},
  issn = {1612-8850, 1612-8869},
  doi = {10.1002/ppap.200900078},
  urldate = {2024-11-22},
  copyright = {http://onlinelibrary.wiley.com/termsAndConditions\#vor},
  langid = {english}
}

@article{slikboer_charge_2017,
  title = {Charge Transfer to a Dielectric Target by Guided Ionization Waves Using Electric Field Measurements},
  author = {Slikboer, Elmar and {Garcia-Caurel}, Enric and Guaitella, Olivier and Sobota, Ana},
  year = {2017},
  month = feb,
  journal = {Plasma Sources Science and Technology},
  volume = {26},
  number = {3},
  pages = {035002},
  issn = {1361-6595},
  doi = {10.1088/1361-6595/aa53fe},
  urldate = {2024-11-17},
  langid = {english}
}

@article{tendero_atmospheric_2006,
  title = {Atmospheric Pressure Plasmas: {{A}} Review},
  shorttitle = {Atmospheric Pressure Plasmas},
  author = {Tendero, Claire and Tixier, Christelle and Tristant, Pascal and Desmaison, Jean and Leprince, Philippe},
  year = {2006},
  month = jan,
  journal = {Spectrochimica Acta Part B: Atomic Spectroscopy},
  volume = {61},
  number = {1},
  pages = {2--30},
  issn = {05848547},
  doi = {10.1016/j.sab.2005.10.003},
  urldate = {2024-11-18},
  copyright = {https://www.elsevier.com/tdm/userlicense/1.0/},
  langid = {english}
}

@article{vanderschans_electric_2017,
  title = {Electric Field Measurements on Plasma Bullets in {{N}}{\textsubscript{2}} Using Four-Wave Mixing},
  author = {Van Der Schans, Marc and B{\"o}hm, Patrick and Teunissen, Jannis and Nijdam, Sander and IJzerman, Wilbert and Czarnetzki, Uwe},
  year = {2017},
  month = oct,
  journal = {Plasma Sources Science and Technology},
  volume = {26},
  number = {11},
  pages = {115006},
  issn = {1361-6595},
  doi = {10.1088/1361-6595/aa9146},
  urldate = {2024-11-17},
  langid = {english}
}

@article{viegas_interaction_2020,
  title = {Interaction of an Atmospheric Pressure Plasma Jet with Grounded and Floating Metallic Targets: Simulations and Experiments},
  shorttitle = {Interaction of an Atmospheric Pressure Plasma Jet with Grounded and Floating Metallic Targets},
  author = {Viegas, Pedro and Hofmans, Marlous and Van Rooij, Olivier and Obrusn{\'i}k, Adam and L M Klarenaar, Bart and Bonaventura, Zdenek and Guaitella, Olivier and Sobota, Ana and Bourdon, Anne},
  year = {2020},
  month = sep,
  journal = {Plasma Sources Science and Technology},
  volume = {29},
  number = {9},
  pages = {095011},
  issn = {0963-0252, 1361-6595},
  doi = {10.1088/1361-6595/aba7ec},
  urldate = {2024-11-17},
  langid = {english}
}

@article{viegas_physics_2022,
  title = {Physics of Plasma Jets and Interaction with Surfaces: Review on Modelling and Experiments},
  author = {Viegas, Pedro and Slikboer, Elmar and Bonaventura, Zdenek and Guaitella, Olivier and Sobota, Ana and Bourdon, Anne},
  year = {2022},
  month = may,
  journal = {Plasma Sources Science and Technology},
  volume = {31},
  number = {5},
  pages = {053001},
  publisher = {IOP Publishing},
  doi = {10.1088/1361-6595/ac61a9}
}

@article{wasik_measurements_1969,
  title = {Measurements of Gaseous Diffusion Coefficients by a Gas Chromatographic Technique},
  author = {Wasik, S.P. and McCulloh, K.E.},
  year = {1969},
  month = mar,
  journal = {Journal of Research of the National Bureau of Standards Section A: Physics and Chemistry},
  volume = {73A},
  number = {2},
  pages = {207},
  issn = {0022-4332},
  doi = {10.6028/jres.073A.018},
  urldate = {2024-12-15},
  langid = {english}
}

@article{weltmann_atmosphericpressure_2010,
  title = {Atmospheric-Pressure Plasma Sources: {{Prospective}} Tools for Plasma Medicine},
  shorttitle = {Atmospheric-Pressure Plasma Sources},
  author = {Weltmann, Klaus Dieter and Kindel, Eckhard and Von Woedtke, Thomas and H{\"a}hnel, Marcel and Stieber, Manfred and Brandenburg, Ronny},
  year = {2010},
  month = apr,
  journal = {Pure and Applied Chemistry},
  volume = {82},
  number = {6},
  pages = {1223--1237},
  issn = {1365-3075, 0033-4545},
  doi = {10.1351/PAC-CON-09-10-35},
  urldate = {2024-11-17},
  langid = {english}
}

@article{winter_atmospheric_2015,
  title = {Atmospheric Pressure Plasma Jets: An Overview of Devices and New Directions},
  shorttitle = {Atmospheric Pressure Plasma Jets},
  author = {Winter, J and Brandenburg, R and Weltmann, K-D},
  year = {2015},
  month = oct,
  journal = {Plasma Sources Science and Technology},
  volume = {24},
  number = {6},
  pages = {064001},
  issn = {0963-0252, 1361-6595},
  doi = {10.1088/0963-0252/24/6/064001},
  urldate = {2024-11-18},
  langid = {english}
}

@article{Xiong_2010,
  title = {On the Velocity Variation in Atmospheric Pressure Plasma Plumes Driven by Positive and Negative Pulses},
  author = {Xiong, Z. and Lu, X. and Xian, Y. and Jiang, Z. and Pan, Y},
  year = {2010},
  month = nov,
  journal = {Journal of Applied Physics},
  volume = {108},
  number = {10},
  eprint = {https://pubs.aip.org/aip/jap/article-pdf/doi/10.1063/1.3511448/14809369/103303{\textbackslash}\_1{\textbackslash}\_online.pdf},
  pages = {103303},
  issn = {0021-8979},
  doi = {10.1063/1.3511448}
}

@article{yan_gas_2017,
  title = {Gas Flow Rate Dependence of the Discharge Characteristics of a Helium Atmospheric Pressure Plasma Jet Interacting with a Substrate},
  author = {Yan, Wen and Economou, Demetre J},
  year = {2017},
  month = oct,
  journal = {Journal of Physics D: Applied Physics},
  volume = {50},
  number = {41},
  pages = {415205},
  issn = {0022-3727, 1361-6463},
  doi = {10.1088/1361-6463/aa8794},
  urldate = {2024-11-17},
  langid = {english}
}

@article{zauner-wieczorek_ion_2022,
  title = {The Ion--Ion Recombination Coefficient {{{\emph{$\alpha$}}}} : {{Comparison}} of Temperature- and Pressure-Dependent Parameterisations for the Troposphere and Stratosphere},
  shorttitle = {The Ion--Ion Recombination Coefficient {{{\emph{$\alpha$}}}}},
  author = {{Zauner-Wieczorek}, Marcel and Curtius, Joachim and K{\"u}rten, Andreas},
  year = {2022},
  month = sep,
  journal = {Atmospheric Chemistry and Physics},
  volume = {22},
  number = {18},
  pages = {12443--12465},
  issn = {1680-7324},
  doi = {10.5194/acp-22-12443-2022},
  urldate = {2024-07-07},
  langid = {english}
}

@article{bruggeman_mass_2010,
	title = {Mass spectrometry study of positive and negative ions in a capacitively coupled atmospheric pressure {RF} excited glow discharge in {He}–water mixtures},
	volume = {43},
	issn = {0022-3727, 1361-6463},
	url = {https://iopscience.iop.org/article/10.1088/0022-3727/43/1/012003},
	doi = {10.1088/0022-3727/43/1/012003},
	number = {1},
	urldate = {2025-03-19},
	journal = {Journal of Physics D: Applied Physics},
	author = {Bruggeman, Peter and Iza, Felipe and Lauwers, Daniël and Gonzalvo, Yolanda Aranda},
	month = jan,
	year = {2010},
	pages = {012003},
}
\newpage

\end{document}